\documentclass{jpp}
\usepackage{graphicx}
\usepackage{bm}
\usepackage[utf8]{inputenc}
\usepackage[T1]{fontenc}
\usepackage{amsmath}
\usepackage{mathtools}
\usepackage{hyperref}

\usepackage{cancel}
\usepackage{subcaption}
\renewcommand{\p}[2]{\ensuremath{\frac{\partial #1}{\partial #2}}}
\newcommand{\tp}[0]{\theta'}
\newcommand{\zp}[0]{\zeta'}

\shorttitle{Surface Current Optimization and Coil-Cutting Algorithms}
\shortauthor{D. Panici et al}

\title{Surface Current Optimization and Coil-Cutting Algorithms for Stage-Two Stellarator Optimization}

\author{D. Panici\aff{1}\corresp{\email{dpanici@princeton.edu}}, R. Conlin\aff{2}, R. Gaur\aff{5}, D. W. Dudt\aff{4}, Y. G.  Elmacioglu\aff{1}, M. Landreman\aff{2}, T. Elder\aff{2},N. Snir\aff{3}, I. Gissis\aff{3}, Y. Nikulshin\aff{3}  \and E. Kolemen\aff{1}\corresp{\email{ekolemen@princeton.edu}}}

\affiliation{\aff{1}Princeton University, Princeton, New Jersey 08544
\aff{2}Institute for Research in Electronics and Applied Physics, University of Maryland, College Park, MD 20742, USA
\aff{3} nT-Tao, Israel
\aff{4} Thea Energy, USA
\aff{5} University of Wisconsin-Madison, Madison, Wisconsin 53706
}

\begin{document}

\maketitle

\begin{abstract}
    Stellarator optimization often takes a two-stage approach, where in the first stage the boundary is varied in order to optimize for some physics metrics, while in the second stage the boundary is kept fixed and coils are sought to generate a magnetic field that can recreate the desired stellarator. Past literature dealing with this stage lacks details on the coil cutting procedure and the mathematical and physical properties of the surface current potential which dictates it. In this work, some basic physical quantities of the surface current and how they relate to the parameters in the current potential are presented, and supported for the first time by explicit mathematical derivations. Additionally, the details of how to account for the presence of an external field in the surface current algorithm are explicitly presented. These relations underpin the procedure of discretizing the surface current into coils. Finally, the conventionally-used algorithm for discretizing the surface current into coils is detailed, along with an example coil optimization for both a modular and a helical coilset. The algorithm is implemented in the \texttt{DESC} code, with both modular and helical coil capabilities, where it is available for use in stellarator coil design.
\end{abstract}

\section{Introduction}
Stellarator coil design comprises the second stage of the conventional two-stage stellarator optimization approach. Stage-one of the optimization utilizes a three-dimensional ideal magnetohydrodynamics (MHD) equilibrium code running in fixed-boundary mode to optimize the shape of the plasma boundary (with the profiles either prescribed or part of the optimization) in order to minimize given physics-based objectives, such as particle confinement (through a proxy like quasisymmetry) and avoidance of rational surfaces in the rotational transform profile. Once a satisfactory fixed boundary equilibrium (as defined by the profiles, enclosed toroidal flux, and the boundary shape) is found, the second stage of the optimization involves finding an external coilset that can provide the necessary enclosed toroidal flux, while also ensuring the plasma boundary is indeed a flux surface (i.e. the total normal magnetic field on that surface is zero) \citep{boozer_stellarator_2000}. It has been observed by Merkel that this second requirement cannot be exactly satisfied \citep{merkel_solution_1987}, so instead a minimum of the integral of the normal field squared is sought, i.e. seeking a coilset which minimizes 
\begin{equation}
    \chi^2_B \coloneqq \int_{S_{plasma}} (\bm{B} \cdot \bm{n})^2 dA,
\end{equation}
where $S_{plasma}$ is the plasma boundary, $\bm{n}$ is the unit surface normal vector and $\bm{B}$ is the total magnetic field (with contributions from the coils external to the plasma and from the plasma field itself). There also has been recent work done on the concept of "single-stage" optimization, where both the plasma boundary and the external coils are optimized simultaneously \citep{henneberg_combined_2021,lee_stellarator_2023,giuliani_single-stage_2022,jorge_single-stage_2023,jorge_simplified_2024}, but this paper will not discuss the single-stage approach.
\par
Stage-two optimization of the coilset is typically approached one of two ways. The first assumes a fixed winding surface upon which a surface current lies, and minimizing the above metric with respect to this current, which we will call the surface current formalism \citep{merkel_solution_1987}. The second is to parameterize a discrete coilset, either filamentary or finite-build, and minimize $\chi^2_B$ with respect to the degrees of freedom of this coilset. The first approach leads to a linear least-squares problem that can be efficiently solved for the global optimum, but is difficult to add other coil objectives to, while the second is a nonlinear minimization problem which can be augmented with other coil complexity objectives and sample a wider range of configurations not restricted to a pre-defined winding surface. Among the codes that perform this second approach are ONSET \citep{drevlak_automated_1998}, COILOPT (though the coils in this code were still restricted to a winding surface) \citep{strickler_designing_2002}, COILOPT++ \citep{gates_recent_2017}, and FOCUS \citep{zhu_new_2018}. Furthermore, the surface current formalism has been shown to be related to the problem of minimizing normal fields using permanent magnet (PM) arrays \citep{helander_stellarators_2020}, and a number of methods and codes have been created to address the PM approach in both continuous and discrete formulations \citep{kaptanoglu_topology_2023,landreman_calculation_2021,hammond_improved_2024,kaptanoglu_greedy_2023, elder_current_2024}.  
\par
While the prior works on the surface current formalism all provide detail on the general algorithm, the details of the coil cutting procedure are not elaborated upon by many of the references, nor are the properties of the surface current potential that inform the coil cutting procedures. This work intends to explain these in a clear way, for ease of introduction to this useful tool of stage-two stellarator optimization.
In~\S\ref{sec:lit-review} we summarize the existing literature on the surface current formalism for coil optimization. We then present the surface current formalism in~\S\ref{sec:surf-current-def}. Section~\S\ref{sec:relations} will present some basic physical quantities in the surface current and how it relates to the parameters in the current potential, supported for the first time by some basic mathematical derivations underpinning the procedure of discretizing the surface current into coils. Then,~\S\ref{sec:coil-cutting} will detail the algorithm for discretizing the surface current into coils that is conventionally used in the community, both for modular and helical coils. Example stage-two coil optimizations with modular and helical coils are then presented with these methods.

\section{Literature Review}\label{sec:lit-review}

% Merkel 1987 (NESCOIL)
The surface current formulation of the stage-two coil optimization problem was first introduced by Rehker and Wobig \citep{s_rehker_stellarator_1973}. Merkel \citep{merkel_solution_1987} extended this formulation to winding surfaces placed arbitrarily far away from the plasma surface with the NESCOIL code. This code has been used to obtain coilsets for optimized stellarators given a fixed winding surface \citep{jr_use_2001,ku_modular_2010,ku_stellarator_2010}, used in-situ along with the stage one optimization to inform coil complexity objectives \citep{zarnstorff_physics_2001,drevlak_optimisation_2018}, or used to create coilsets as starting points for further refinement with nonlinear coil optimization codes \citep{drevlak_esl_2013,zarnstorff_physics_2001}.
% Pomphrey 2001 (Truncated SVD)
Later, NESCOIL was modified to include a truncated singular value decomposition (TSVD) in a code titled NESVD which attempts to better regularize the problem \citep{pomphrey_innovations_2001}.
% Landreman 2017
More recently, the algorithm was modified to include a Tikhinov-like regularization based off of the surface current density magnitude \citep{landreman_improved_2017} which was shown to yield much better behaved results than the earlier formulations. This algorithm was implemented in FORTRAN in the original publication, and as part of this work has been re-implemented inside of the DESC stellarator optimization framework \citep{dudt_desc_2020,conlin_desc_2023,dudt_desc_2023,panici_desc_2023}. The authors note that prior to this work, the \texttt{REGCOIL} algorithm was also re-implemented in Python and augmented with a coil-coil force metric by Robin and Volpe \citep{robin_minimization_2022}. Boozer \citep{boozer_interaction_2021} also independently analyzed the form of the magnetic forces on a surface current. Fu recently described a more general surface current optimization algorithm, \texttt{QUADCOIL}, which can handle linear and quadratic constraints and objectives of the current \citep{fu_global_2024}. Elder has also written on coil-cutting algorithms for discretizing sheet currents \citep{elder_three-dimensional_2024}.

\section{Surface Current Formalism}\label{sec:surf-current-def}

Throughout this work, the primed angular coordinates $(\tp,\zp)$ will be used to refer to the winding surface coordinates, while the unprimed angles $(\theta, \zeta)$ will refer to the plasma surface coordinates. In the surface current density formulation introduced by Merkel and used in NESCOIL and \texttt{REGCOIL}\citep{merkel_solution_1987,landreman_improved_2017}, the functional form of the surface current density is:

\begin{equation}\label{eq:surface_current_density_def}
    \bm{K} =\bm{n} \times \nabla \Phi(\tp,\zp)~~~\text{[A/m]},
\end{equation}

where $\bm{n} = \frac{\bm{e}_{\tp} \times \bm{e}_{\zp}}{|\bm{e}_{\tp} \times \bm{e}_{\zp}|}$ is the unit surface normal vector to the winding surface and $\bm{e}_x = \frac{\partial \bm{r}}{\partial x}$ are the covariant basis vectors \citep{dhaeseleer_flux_2012}, where $x=\{\tp,\zp\}$. This form follows from the need for the surface current on the winding surface to be divergence-free ($\nabla \cdot \bm{K} = 0$ ) and that the current lie entirely on the winding surface ($\bm{K} \cdot \bm{n} = 0$) \citep{boozer_optimization_2000, boozer_non-axisymmetric_2015}. The current potential $\Phi$  is given by a single-valued part $\Phi_{SV}(\tp,\zp)$ and a secular part ($I$ in $\theta'$ and $G$ in $\zeta'$).
\begin{equation}\label{eq:curr_pot}
    \Phi(\theta',\zeta') = \Phi_{SV}(\theta',\zeta') + \frac{G\zeta'}{2\pi} + \frac{I\theta'}{2\pi} ~~~\text{[A]},
\end{equation}

where the secular parts carry a physical meaning: $I$ is the net toroidal current (in Amperes) flowing on the winding surface, while $G$ is the net poloidal current flowing on the winding surface. These relations are derived for an arbitrary nonaxisymmetric surface in Section \ref{sec:relations} and Appendix \ref{app:G_derivation}. The currents on the winding surface flow along contours of constant current potential, as $\bm{K} \cdot \nabla \Phi = 0$, which naturally leads to contours of constant $\Phi$ being used to discretize the current into coils, which will be elaborated upon in Section \ref{sec:coil-cutting}.

\section{Relations of the Surface Current Density to Net Current and Equilibrium Fields} \label{sec:relations}

Consider a toroidal plasma equilibrium and a set of coils external to that equilibrium. The net poloidal current required to be carried by the external coils can be found by considering Ampere's law and integrating over the area inside the torus hole. This then simplifies to a line integral of the covariant toroidal field around one toroidal transit, yielding the relation between the net current in the coils linking the equilibrium poloidally and the equilibrium magnetic field. Taking $A$ as the surface inside the torus hole bounded at the edges by the inboard side of the equilibrium last closed flux surface, and $\partial A$ as the bounding curve of that surface:

\begin{align}
    \iint_A \nabla \times \bm{B} \cdot d\bm{A} &=  \iint_A \mu_0 \bm{J} \cdot d\bm{A}\\ 
    &\text{Using Stoke's theorem on the left-hand side:}\notag\\
    \oint_{\partial A} \bm{B} \cdot d\bm{l} &= \mu_0 I_{pol} 
\end{align}
Where $I_{pol}$ is the net poloidal current flowing through the hole of the plasma torus. Here we neglect any surface currents that may be on the plasma surface.  Using the fact that the only currents inside the torus hole are the coil currents, we find that $I_{pol}$ for a set of coils is equal to:
\begin{align}\label{eq:coil-currents}
     \oint_{\partial A} \bm{B} \cdot d\bm{l}&= \mu_0 I_{pol}= \mu_0 \sum_{i=1}^{N_{coils}} I_{i}^{coil} N_i^{poloidal~ transits}\\
     &\text{$\bm{B} \cdot d\bm{l} = \bm{B}\cdot \frac{\partial \bm{r}}{\partial \zeta} d\zeta = B_{\zeta} d\zeta$, LHS becomes}\\
     \Aboxed{N_{FP}\int_{0}^{2\pi/N_{FP}} B_{\zeta}d\zeta &= \mu_0 I_{pol} = \mu_0 \sum_{i=1}^{N_{coils}} I_{i}^{coil} N_i^{poloidal~transits} }
\end{align}

\begin{figure}
    \centering
    \includegraphics[keepaspectratio,width=2in]{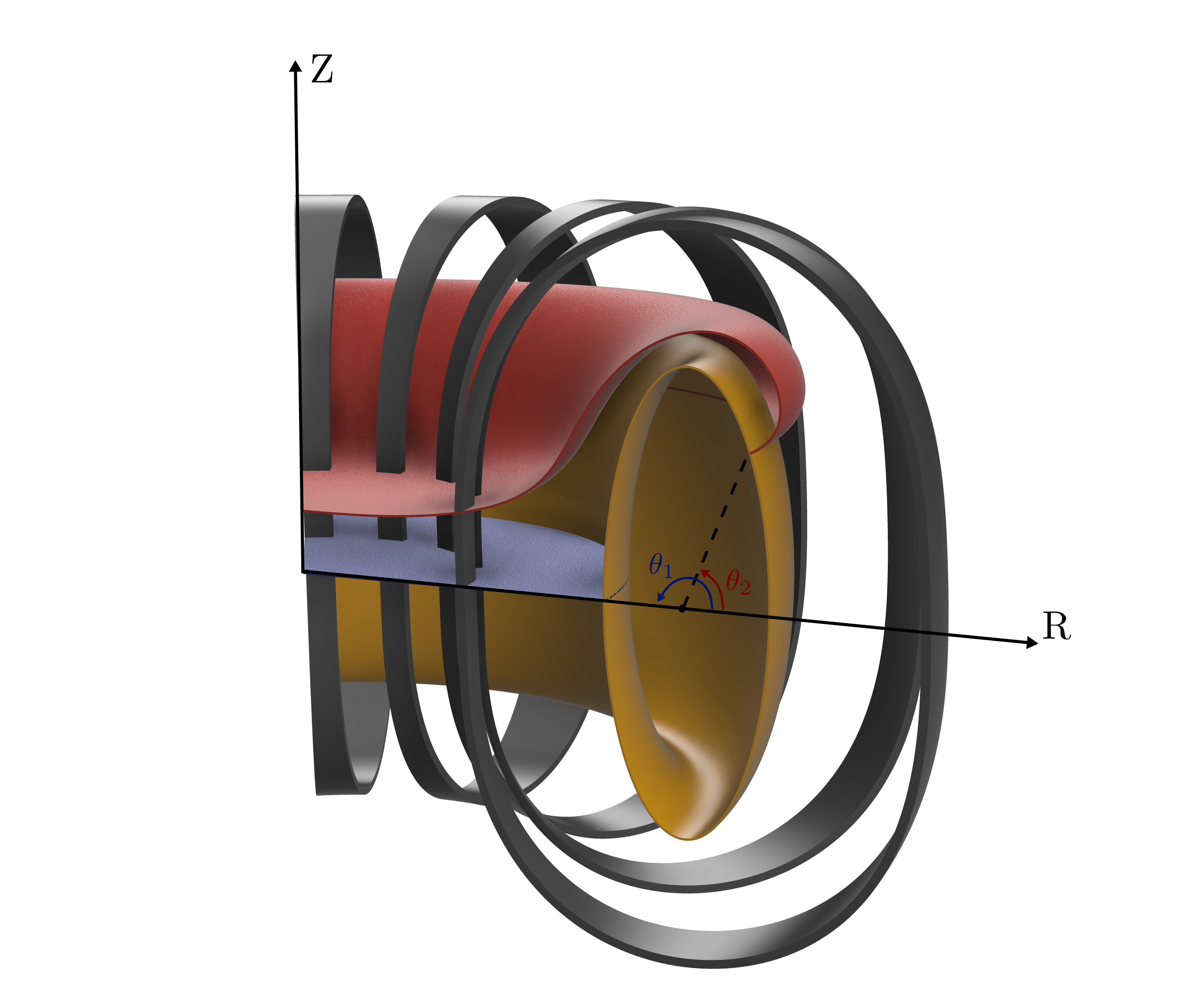}
    \caption{Schematic of two different integration surfaces leading to two different line integrals for the left side of Eq. \eqref{eq:coil-currents}. The blue surface results in an inboard contour at $\theta_1$, while the red surface results in an outboard contour at $\theta_2$. Both are valid and satisfy Eq. \eqref{eq:coil-currents}, as both surfaces enclose the same net poloidal current (represented in this case by the coils in gray).}
    \label{fig:surface_integral_thetas}
\end{figure}

Where $\bm{B}$ is the equilibrium magnetic field, $I_i^{coil}$ is the current flowing through the $i^{\text{th}}$ coil, and $N_i^{poloidal~transits}$ is the number of poloidal transits made by the $i^{\text{th}}$ coil (this is the linking number between the coils and the torus axis, which is one for modular coils, possibly greater for helical coils, and zero for windowpane coils). The RHS is thus the total current linking the plasma poloidally (i.e. the currents that go around the plasma poloidally from the coils). A schematic view of this is shown in Figure \ref{fig:surface_integral_thetas}, where the blue and red surfaces are two valid choices of the surface $A$, which would result in line integrals at constant theta taken along the inboard and outboard side of the plasma, respectively. These surfaces both enclose the same net poloidal current, and so both satisfy Eq. \eqref{eq:coil-currents}. This is useful as when performing these integrals over the equilibrium field to find the required net poloidal current of the coils, or when identifying the net poloidal current associated with a given external field (discussed in Section \ref{sec:relations}), no heed needs to be paid as to whether the integral \eqref{eq:coil-currents} is taken along the inboard or outboard side, only that it be along a constant theta curve on the boundary.

% the surface $A$ need not be bounded by the inboard side of the plasma, the requirements on $A$ for the above equations to hold are that 
% \begin{itemize}
%     \item the surface $A$ only intersect each coil (or each poloidal turn of the coil, if it is a helical coil) once (i.e. so that the net poloidal current through the surface from each coil is nonzero and the entire contribution to the integral is counted)
%     \item the surface either does not intersect the plasma past the inboard side, or intersects through the inboard and out the outboard side of the plasma.
% \end{itemize} This is because even if the surface contains the plasma, the poloidal currents in the plasma will both flow into and then out of the surface, thus contributing zero current to the RHS of Eq. \eqref{eq:coil-currents}.

% \textcolor{red}{Picture here would help (and probably a calculation to double check this for a rotating ellipse or something)}

% \par
% Consider instead of an external coilset that there is a toroidal winding surface upon which a surface current lies, as described by Eq. \eqref{eq:surface_current_density_def}. The previous points hold, except now the integration surface being considered must intersect every current-carrying contour that closes poloidally on the winding surface, which means that the surface must intersect the entire toroidal circumference of the winding surface.\\
\par
The above equations are related to the form of the surface current density used as follows. First, the surface current's components are defined as:

\begin{align}
    \bm{K} &=\bm{n} \times  \nabla \Phi = \bm{n} \times  \nabla \left( \Phi_{SV}(\theta',\zeta') + \frac{G\zeta'}{2\pi} + \frac{I\theta'}{2\pi} \right) \\
    &= K^{\theta} \bm{e}_{\theta'} + K^{\zeta} \bm{e}_{\zeta'},
\end{align}

where
\begin{align}\label{eq:surf-curr-components}
    K^\theta &= -\frac{1}{|\bm{N}|}\left(\p{\Phi_{SV}}{\zeta'} + \frac{G}{2\pi}\right)\\
    K^\zeta &= \frac{1}{|\bm{N}|}\left(\p{\Phi_{SV}}{\theta'} + \frac{I}{2\pi}\right),
\end{align}

are the contravariant components of the surface current density.

Next, the parameters $I$ and $G$ are connected to their physical meanings. The net current linking the winding surface and the plasma poloidally (i.e. the net poloidal current on the winding surface, $I_{pol}$) is related to the magnetic field inside through  Eq.\eqref{eq:coil-currents}. $I_{pol}$ is also equal to the negative (due to the sign of $K^\theta$ in Eq. \eqref{eq:surf-curr-components}) of the secular term $G$, while the net current flowing toroidally around the winding surface $I_{tor}$ is equal to the (positive) secular term $I$:
\begin{align}
    I_{pol} &=-G\\
    I_{tor} &= I,
\end{align}

Further, the current passing between two constant current potential contours on the winding surface is simply equal to the difference in the current potential $\Phi$ of those two contours:

\begin{align}
    I_{\Delta } &= \Phi_1 - \Phi_0,
\end{align}

where $I_{\Delta}$ is the current flowing between the two contours and $\Phi_1$ and $\Phi_0$ are the value of the current potential along the two given contours. Each of these statements is derived in Appendix \ref{app:G_derivation}, for an arbitrary toroidal winding surface. These relations physically motivate the algorithms for discretization of the surface current into coils detailed in Section \ref{sec:cutting_helical_coils}, and thus understanding where they come from can lend insight into the discretization procedure.

\section{\texttt{NESCOIL} and \texttt{REGCOIL} Algorithm}\label{sec:nescoil-and-regcoil}

The problem solved by the \texttt{NESCOIL} code, the original formulation of the surface current approach to stage-two coil optimization, is to minimize the quadratic flux through the plasma surface with respect to the degrees of freedom in the surface current on the winding surface (which is the single-valued part $\Phi_{SV}$, as $I$ and $G$ are determined by desired topology and the equilibrium field, respectively):

\begin{equation}
    \min_{\Phi_{SV}}~ \chi^2_B,
\end{equation}

with 
\begin{equation}
    \chi^2_B \coloneqq \int_{S_{plasma}} (B_n)^2 dA,
\end{equation}

where 

\begin{equation}
    B_n = B_n^{SV}\{\Phi_{SV}\} + B_n^{GI} +B_n^{plasma} + B_n^{ext},
\end{equation}

is the total magnetic field normal to the plasma surface due to the individual components from the single-valued surface current potential $B_n^{SV}$, the secular part of the surface current potential $B_n^{GI}$, the plasma currents $B_n^{plasma}$ and any external fields $B_n^{ext}$ (a toroidal field (TF) coilset, PM, or any other coilset besides the winding surface, see e.g. \citep{pomphrey_innovations_2001, jr_use_2001}). Given that the plasma surface and winding surface geometries are fixed, it can be shown \citep{landreman_improved_2017} that the normal magnetic field component due the single-valued current potential is linear in $\Phi_{SV}$. The other components are known given a plasma and any external coilsets, thus the problem essentially is to find the single-valued current potential $\Phi_{SV}$ such that the normal field it creates, $B_n^{SV}$, cancels out (as much as possible) the combined contributions from the other components, to attempt to find the closest solution to $B_n=0$.

The single-valued surface current potential is typically represented by a periodic double Fourier basis in $(\theta',\zeta')$:

\begin{equation}
    \Phi_{SV}(\tp,\zp) = \sum_{m,n} \Phi_{SV}^{mn} \mathcal{F}_m(\tp)\mathcal{F}_{nN_{FP}}(\zp)   ,  
\end{equation}

where the Fourier series is defined as:

\begin{align}
    \mathcal{F}_n(x) = \begin{cases} cos(|n|x) \text{for } n\geq 0 \\
    sin(|n|x) \text{for } n < 0 
    \end{cases}
\end{align}

If both the winding surface and the equilibrium are stellarator-symmetric, then $\Phi_{SV}$ itself is "sin" symmetric, that is if written out in the double angle Fourier form, it would only have terms like $sin(m\tp - n\zp)$ \citep{merkel_solution_1987}. 
\par
The quadratic flux cost $\chi^2_B$ in this minimization problem is thus quadratic in the unknown $\Phi_{SV}^{mn}$ \citep{landreman_improved_2017}, so it has the form of a linear least-squares problem, and has a unique global solution which can be found by solving a single linear system. 
However, while having a unique global solution, the problem can still be ill-conditioned \citep{landreman_improved_2017}. The \texttt{REGCOIL} algorithm adds a regularization term based off of the surface current density magnitude on the winding surface $S_{winding}$:

\begin{equation}
    \chi^2_K = \int_{S_{winding}} |\bm{K}|^2 dA',
\end{equation}

yielding the regularized minimization problem:

\begin{equation}
    \min_{\Phi_{SV}^{mn}} ~\chi^2_B + \lambda \chi^2_K,
\end{equation}

where $\lambda$ is the regularization parameter, which can be varied to control the level of regularization in the problem. A small $\lambda$ will tend to yield a solution with lower quadratic flux but higher current densities, while a larger $\lambda$ will tend to increase the quadratic flux on the plasma surface but decrease the surface current density magnitude on the winding surface \citep{landreman_improved_2017}. \\
Both terms in this minimization problem are quadratic in the unknown $\Phi_{SV}^{mn}$ \citep{landreman_improved_2017}, so it still has the form of a linear least-squares problem, and thus has a unique global solution for a given regularization parameter value $\lambda$. 

\section{Discretizing Surface Currents into Coils}\label{sec:coil-cutting}

% here talk about div theorem and how it makes sense to give a single contour the current flowing btwn halfway to its neighbors?
Once the surface current which minimizes $B_n$ on the plasma surface is found, the continuous surface current must be discretized into filamentary coils. An algorithm conventionally used in the community (though only mentioned in any form by a few references \citep{boozer_optimization_2000, pomphrey_innovations_2001, jr_use_2001, drevlak_automated_1998, elder_three-dimensional_2024}) is to pick $N_c$ equally spaced current potential contours and use them as the coils, with the currents in each of them being equal (a logical choice of contours and currents in each coil that was motivated in Section \ref{sec:relations}). 
\par
Work performed by Miner during the NCSX coil design leveraged equally-spaced contours as an initial guess for further global optimization with a genetic algorithm \citep{jr_use_2001}. However, they also solved a sub-problem to minimize the quadratic flux with respect to the coil currents after selecting candidate coil contours.

\subsection{Compensating for External TF Coil Poloidal Currents}

% cite Pomphrey and Miner for this

The surface current potential, as previously mentioned, is given by:

\begin{equation}\label{eq:currpot}
        \Phi(\theta',\zeta') = \Phi_{SV}(\theta',\zeta') + \frac{G\zeta'}{2\pi} + \frac{I\theta'}{2\pi},
\end{equation}

where $(\theta',\zeta')$ are poloidal and toroidal angles on the winding surface, $G$ is total current linking the coil surface poloidally (which is zero for windowpane coils, and for modular or helical coils is determined entirely by the target plasma equilibrium magnetic field), and $I$ is the current linking the coil surface toroidally.
\par
In typical usages of \texttt{REGCOIL}, it is desired to find a coilset on the winding surface which completely provides the fields needed to get as close to $B_n=0$ as possible (by least-squares minimization) on the provided last-closed-flux-surface of the plasma equilibrium.
It is also of interest to find a coilset on a given winding surface which is supplemented by a toroidal field coilset (or some other coilset) other than the provided winding surface, where this other coilset provides some of the TF for the magnetic field, or perhaps some other shaping field (similar ideas have been explored in the design of NCSX \citep{pomphrey_innovations_2001, jr_use_2001}, though the following necessary modification to the current potential has not been stated explicitly in past works). The typical way $G$ is chosen for a modular or helical coilset assumes that the current potential on the winding surface must  provide the \textit{entirety} of the net poloidal current dictated by Ampere's law. However, if there exists another coilset external to the plasma, but not on the winding surface being considered, that coilset may also have a net current linking the plasma poloidally. In this case, one must subtract this current from the $G$ in Eq. \ref{eq:curr_pot} (defining $G_{ext}$ as the net poloidal current contained in coils not on the winding surface of interest):

\begin{equation}
        \Phi(\theta',\zeta') = \Phi_{SV}(\theta',\zeta') + \frac{(G-G_{ext})\zeta'}{2\pi} + \frac{I\theta'}{2\pi},
\end{equation}

So that Ampere's law is still respected, and thus reducing the amount of net poloidal current required to be carried in the \texttt{REGCOIL} solution. Some example uses include finding a shaping coilset while having simpler TF coils provide the bulk of the toroidal magnetic flux, or searching for a shaping coilset with a supplementary set of simple helical coils surrounding the plasma, in an attempt to simplify the overall design \citep{pomphrey_innovations_2001, ku_nonaxisymmetric_2009}.

\subsubsection{Finding the Net Poloidal Current due to an External Field or Coilset}

In the case given above, we must calculate what the $G_{ext}$ is so that it can be subtracted from the total $G$. Given a coilset, one can determine the net poloidal current simply through adding up the currents of each coil, multiplied by their linking number with the plasma (the number of times they encircle the plasma poloidally, this is one for modular coils, zero for saddle, and may be greater than one for helical coils.). However, one must be careful to understand the physical direction of the current, in order to correctly choose the sign of the external net poloidal current  with respect to the net poloidal current of the winding surface. As a check, given any magnetic field, one can determine the corresponding external poloidal linking current associated with that field by simply taking the toroidal field generated by the coils, $B_{\zeta}^{ext}$, and integrating it along a curve on the plasma boundary, at constant poloidal angle $\theta$:

\begin{align}
    \frac{1}{\mu_0}\oint_{\partial A} \bm{B}_{ext}\cdot d\bm{l} =\iint_{A} \bm{J}_{ext} \cdot d\bm{A} &=: G_{ext}\\
    \frac{1}{\mu_0}\int_{0}^{2\pi} B_{\zeta}^{ext}d\zeta &= G_{ext}
\end{align}

Where $B_{\zeta}^{ext} = \bm{B}_{ext} \cdot \bm{e}_{\zeta}$ and $\bm{J}_{ext}$ is the current density responsible for generating the external magnetic field. 
% So, with this, one may employ the \texttt{REGCOIL} algorithm to find surface currents minimizing the normal magnetic field subject to any external magnetic beyond the solution sought on the desired winding surface. This functionality has been implemented in the \texttt{DESC} version of \texttt{REGCOIL}.

\subsection{Surface Current Contour Topology} 

In this section, the connection of the constant current potential contour topology to the parameters of the surface current is detailed. Consider the task of finding a surface current that corresponds to helical coils. While $G$ is a value which is determined entirely by the MHD equilibrium solution (subject to possibly the external coil poloidal current subtraction), one must specify the net current linking the toroidally, $I$. This parameter determines the topology of the total current potential contours (and consequently, the types of coils that they correspond to). A value of $I=0$ with $G\neq 0$ will yield modular coils (which only link the plasma poloidally), a value of $G=0$ with $I\neq0$ will yield coils similar to vertical field coils (which only link the plasma toroidally), while a value of $I=0$ with $G=0$ will only yield saddle/windowpane coils (as there can be no net poloidal or toroidal current in the coils, so they cannot enclose the plasma at all). A non-zero value of both $I$ and $G$ will tend to yield current potential contours which can close helically. However, conditions exist on $I$ and $G$ for these contours to close with an integer periodicity (i.e. helical coils which terminate after a single toroidal transit). Note that in this section we are assuming that windowpane coils are not being sought, which correspond to both $I$ and $G$ being set to zero. \footnote{Elder has written on potential methods to cut coils when windowpanes are present \citep{elder_three-dimensional_2024}}. Further, we are assuming that there are no local extrema in the current potential $\Phi$ so that contours do not circle themselves, which tends to occur when the winding surface is far from the plasma or the regularization parameter is small, so that there exist points on the surface where $\Phi_{SV}$ becomes comparable in size to the secular terms, leading to $|\nabla \Phi|=0$.

Consider the closure conditions for a contour of constant current potential $\Phi$ on a winding surface with field periodicity $N_{FP}$ to return to the same poloidal angle  after some integer $p$ toroidal field period transits (e.g. if $p=1$, it closes back on itself after $N_{FP}$ field period transits, a usually desirable property for helical coils, which would have the same field periodicity as the underlying surface and equilibrium):
\begin{align}
    \Phi(\tp_0, \zp_0) &= \Phi(\tp_f, \zp_0 + 2\pi p/N_{FP})\\
    \text{where } mod(\tp_f, 2\pi) &= \tp_0
\end{align}
Plugging in $\Phi$ from eq. \eqref{eq:currpot} which has both periodic and secular components in $\tp$ and $\zp$ gives:

\begin{align}
    \Phi_{SV}(\tp_0,\zp_0) + \frac{(G-G_{ext})\zeta_0}{2\pi} + \frac{I\tp_0}{2\pi} &= \Phi_{SV}(\tp_f,\zp_0+2\pi p/N_{FP})\notag\\
    &+ \frac{(G-G_{ext}) (\zp_0+2\pi p/N_{FP})}{2\pi } + \frac{I\tp_f}{2\pi}.
\end{align}

\begin{figure}
    \centering
    \includegraphics[keepaspectratio, width=4 in]{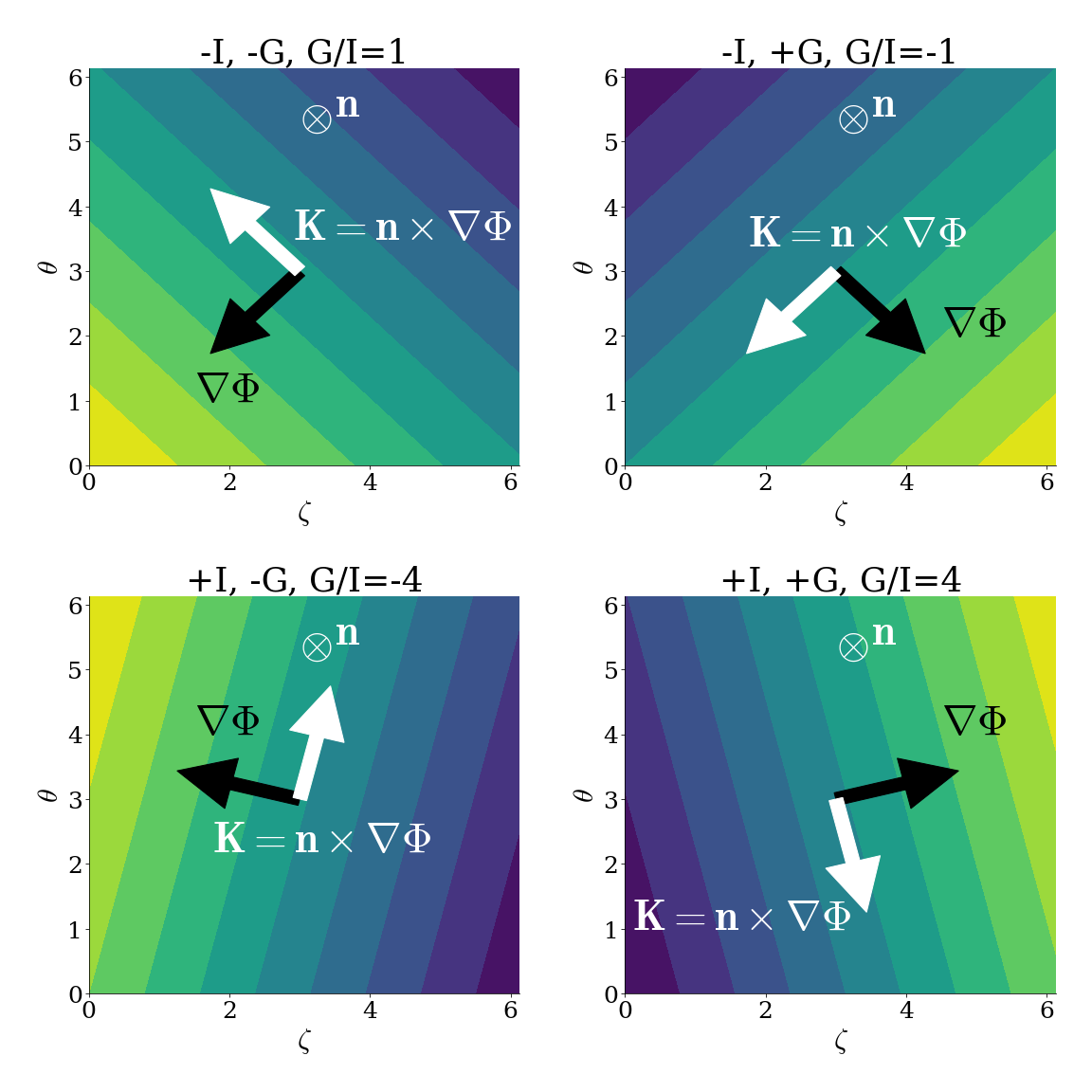}
    \caption{Contour plots of the current potential $\Phi$ and the directions of the gradient of the current potential $\nabla \Phi$, the surface current $\mathbf{K}$ and the unit surface normal vector $\mathbf{n}$, for different combinations of signs of $I$ and $G$ (assuming for simplicity that $G_{ext}=0$, $N_{FP}=1$, $p=1$ and $\Phi_{SV}=0$). Note that only the ratio $G/I = \frac{q}{p}$ is what determines the topology of the contours (and the subsequent coils created), while the sign of both $I$ and $G$ (or equivalently, $p$ and $q$) matter when determining the direction in which current flows.}
    \label{fig:current_helicity}
\end{figure}

 $\Phi_{SV}$ is periodic in $\tp$ (with period $2\pi$) and $\zp$ (with period $2\pi/N_{FP}$), and since part of the closure condition is that $mod(\tp_f, 2\pi) = \tp_0 $, the single-valued parts of this equation on both sides are equal and thus cancel out. This leaves us with the secular parts, which can be rearranged for $\tp_f$ to yield a condition on $G$ and $I$:

\begin{align}\label{eq:helicity_def}
     \frac{(G-G_{ext})\zp_0}{2\pi} + \frac{I\tp_0}{2\pi} &= \frac{(G-G_{ext})(\zp_0+2\pi p/N_{FP}) }{ 2\pi } + \frac{I\tp_f}{2\pi}\\
     (\tp_f - \tp_0)/2\pi &= - \frac{p(G - G_{ext})}{I N_{FP}} =: \bar{q} = \frac{q}{N_{FP}},
      % (\tp_f - \tp_0)/2\pi &= - \frac{(G - G_{ext})}{I N_{FP}} =: \frac{\bar{q}}{p} = \frac{q}{p N_{FP}},
\end{align}
where $q$ is defined as the total number of poloidal transits the contour makes before coming back to itself, $\bar{q}\coloneqq\frac{q}{N_{FP}}$ is the number of poloidal transits per field period that the contour makes, and $p$, as previously defined, can also be identified as the number of full toroidal transits that the contour makes before returning to itself (which we will take equal to one for the remainder of this paragraph as we consider only contours which close after a single full toroidal transit). 
Since we have the closure condition  $mod(\tp_f, 2\pi) = \tp_0 $, we necessarily have that $(\tp_f - \tp_0)/2\pi\in \mathbb{Z}$, hence $\bar{q}$ is also an integer. The integer sign simply determines the direction of the contour helicity on the torus, i.e. if positive then $\tp_f>\tp_0$ so the contours go from bottom left to upper right if plotted on $(\zp, \tp)$, like a line that is $\tp = +\zeta$, and vice-versa. Figure \ref{fig:current_helicity} illustrates this for different combinations of signs of $I$ and $G-G_{ext}$. This in turn sets a condition on our tunable parameter $I$, namely that for a given $G$ and $N_{FP}$ (which are determined by the equilibrium), and for a given externally-generated current $G_{ext}$, if we want helically closed contours (specifically, contours which are helical and close after a single full toroidal transit, so $p=1$)
we must choose $I$ such that $\frac{\bar{q}}{p} =\bar{q}= -\frac{G - G_{ext}}{I N_{FP}} \in \mathbb{Z}$ i.e. so that $\bar{q}$ is an integer, corresponding to the number of poloidal transits per field period desired for the coils. Concretely, if we desire helical coils which wrap around the winding surface poloidally $2$ times per field period (so $|\bar{q}|=2$), and wrap in the direction of increasing $\tp$ (so $sign(\bar{q})>0$), we would choose $I$ s.t. $-\frac{G - G_{ext}}{I N_{FP}} = \bar{q}= 2$, leading us to set $I = -\frac{G - G_{ext}}{(2) N_{FP}}$. \par
It is interesting to note that the ratio between the toroidal and poloidal currents in the surface current can also be defined in terms of the language of torus knots and unknots. Following the notation of \citep{oberti_influence_2019}, a torus knot/unknot is a symmetric, closed curve on a torus, denoted $\mathcal{T}_{p,q}$ which wraps around the torus $p$ times toroidally and $q$ times poloidally before returning to its initial point (or "biting its own tail" ). A torus knot is given by taking $p>1$ and $q>1$, with $p,q$ being co-prime integers, while a torus knot with either $p=1$ or $q=1$ is topologically equivalent to a torus unknot (essentially, a curve on the torus which does not knot itself, and can be smoothly deformed to a circle). A knot/unknot's topology is determined by the ratio of $q$ to $p$, called the $winding~number:=q/p$. In the prior paragraph, we required that our helical contours close on themselves after a single full transit, so we effectively fixed $p=1$, and the contours could be considered knots with topology $\mathcal{T}_{1,\bar{q}N_{FP}}$. The $winding~ number$ is $(G-G_{ext})/I = \frac{q}{p} = \frac{G-G_{ext}}{I}$. If we assume $N_{FP}=1$ and $G_{ext}=0$, then we see that the ratio $G/I$ becomes the $winding~number$.
\par
\begin{figure}
  \begin{subfigure}[t]{0.5\textwidth}
    \includegraphics[height=2.5in,keepaspectratio]{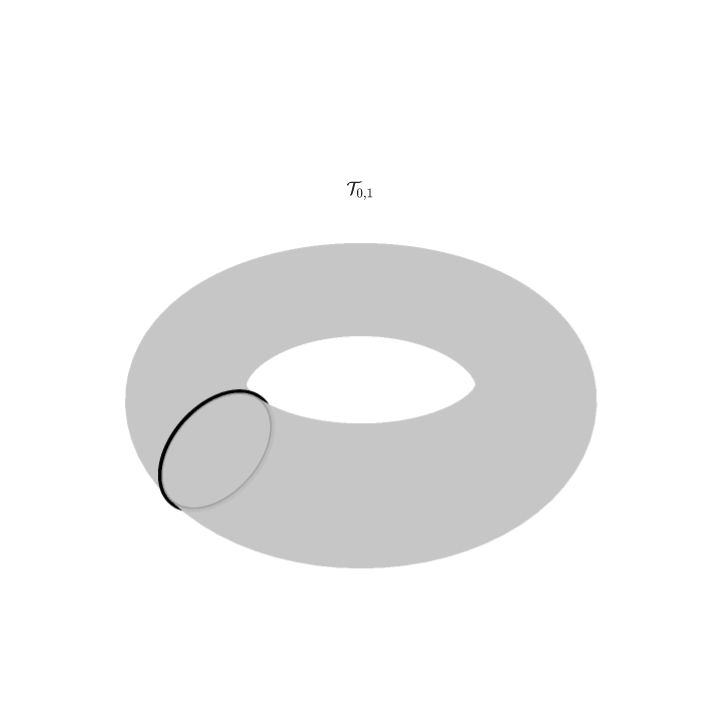}
    \caption{$\mathcal{T}_{0,1}$ unknot, corresponding to a modular coil, where $I=0$ and $G\neq0$}
  \end{subfigure}
  \begin{subfigure}[t]{0.5\textwidth}
    \centering
    \includegraphics[height=2.5in,keepaspectratio]{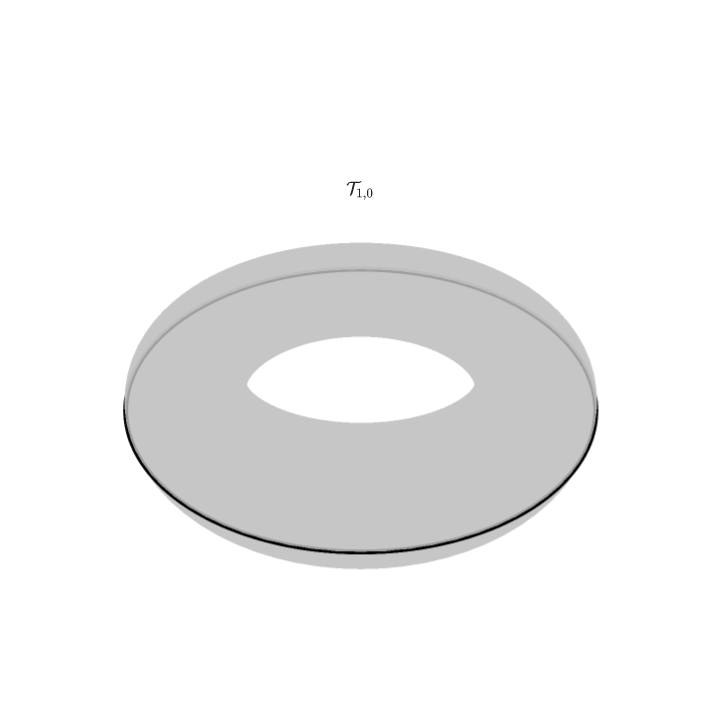}
    \caption{$\mathcal{T}_{1,0}$ unknot, corresponding to a poloidal field coil, where $I\neq0$ and $G=0$.}
  \end{subfigure}
\begin{subfigure}[t]{0.5\textwidth}
    \centering
    \includegraphics[height=2.5in,keepaspectratio]{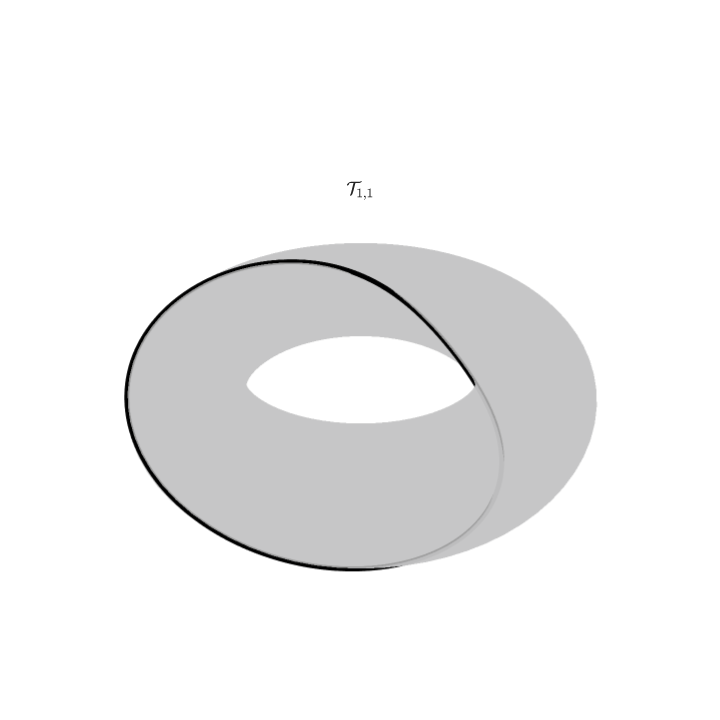}
    \caption{$\mathcal{T}_{1,1}$ unknot, corresponding to a helical coil, where $I=G\neq0$ }
  \end{subfigure}
  \begin{subfigure}[t]{0.5\textwidth}
    \centering
    \includegraphics[height=2.5in,keepaspectratio]{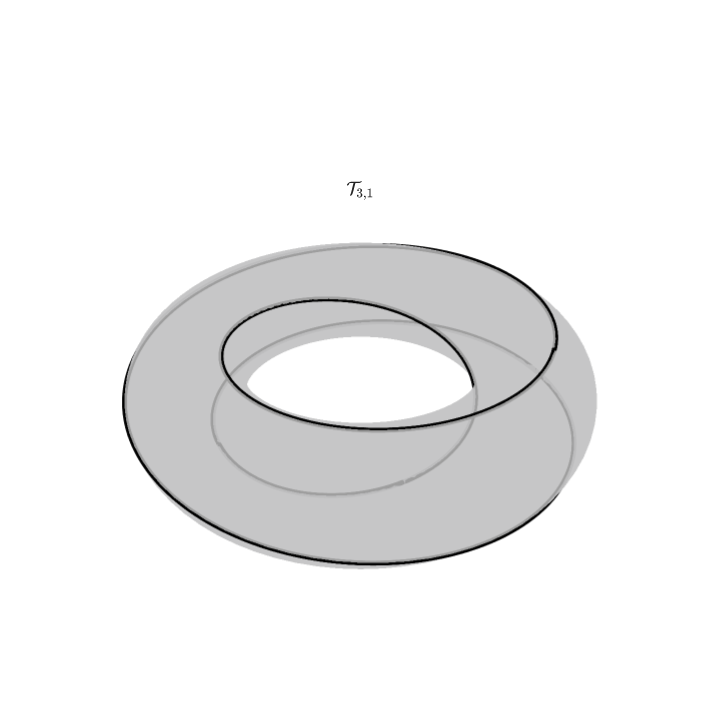}
    \caption{$\mathcal{T}_{3,1}$ unknot, corresponding to an umbilic helical coil, where $I\neq0$, $G\neq0$ and $G/I=3$ }
  \end{subfigure}
  
  \caption{Diagrams of different types of coils and their corresponding knot/unknot.}
  \label{fig:coil_knots}
\end{figure}

If the requirement that the helical coils close on themselves after one toroidal transit is relaxed ($p>1)$, one may find more classes of helical coils. For example, $p=3$ and $q=1$ corresponds to a helical curve which closes on itself after making three complete toroidal transits, while making only one single poloidal transit about the torus. This type of coil may be interesting for umbilic-type stellarator concepts \citep{gaur_omnigenous_2025}, where the stellarator may be thought of as having a fractional $N_{FP}=1/3$, and possibly sharp edges, similar to the "umbilic bracelet" geometric object coined in a work by \citet{zeeman_umbilic_1976} (Here by fractional field periods, we mean that the geometry repeats itself every $2\pi/N{FP}$ in $\zeta$, thus it takes $1/N_{FP}$ toroidal transits for a curve to return to the same point). Figure \ref{fig:coil_knots} shows examples of different classes of helical curves for $N_{FP}=1~G_{ext}=0$, showing modular, toroidal, helical, and umbilic-type curves, with their $G/I$ ratio and corresponding torus unknot analogs.
\par
\begin{figure}
    \centering
    \includegraphics[keepaspectratio, width=3in]{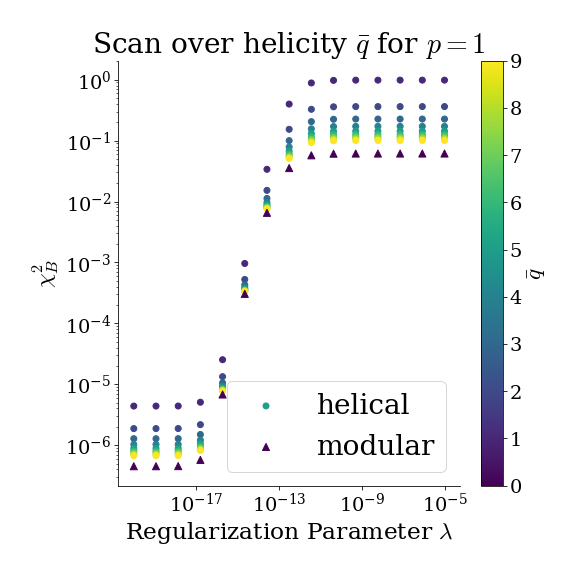}
    \caption{Plot of $\chi^2_B$ versus the regularization parameter $\lambda$ for $p=1$ and a scan over $\bar{q}$, where it can be seen that as $\bar{q}$ is increased, the values tend towards the modular ($p=0$ or equivalently $\bar{q}\rightarrow\infty$) values.}
    \label{fig:helicity_scan}
\end{figure}

Generalizing the topology of coils in the previous paragraphs to coils which close after arbitrary integers $q$ and $p$ transits poloidally and toroidally, respectively, about a torus of field periodicity $N_{FP}$ (instead of requiring the total poloidal transits to be a multiple of $N_{FP}$ through $\bar{q}$), one finds that the required toroidal current is $I = - \frac{p(G - G_{ext})}{q}$. It is satisfying to note that modular coils are obtained in the limit of setting $p \rightarrow 0$ (or $q\rightarrow \infty)$, as then $I \rightarrow 0$ and topologically the curve's ratio of completed toroidal transits per poloidal transit tends to zero. This can be seen in Figure \ref{fig:helicity_scan} where the \texttt{DESC} \texttt{REGCOIL} algorithm is ran for the precise QA \citep{landreman_magnetic_2022} equilibrium for a scan of $\bar{q}$ and $\lambda$ at fixed $p=1$, with a winding surface of a constant offset of 0.2m from the plasma boundary. As the $\bar{q}$ is increased for fixed $p=1$, the $\chi^2_B$ at each regularization parameter value tends towards the modular ($p=0$ or equivalently $\bar{q}\rightarrow\infty$) value. However, for the remainder of this work pertaining to coil cutting, we will only focus on either modular coils $\mathcal{T}_{0,1}$ or helical coils which close after one full toroidal transit which make $\bar{q}=1$ poloidal transits per field period, $\mathcal{T}_{1,q}=\mathcal{T}_{1,\bar{q}N_{FP}}$.

\subsection{Cutting Surface Current into Coils} \label{sec:cutting_helical_coils}
Once a current potential which minimizes the normal field on the plasma surface is found, one needs to  discretize the resulting surface current density distribution into a finite number $N_{coils}$ of filamentary coils. This process can be thought of as having three steps: 
\begin{enumerate}
    \item Finding the coordinates $N_{coils}$ contours of constant current potential on the winding surface, which will be the locations of the coils on the winding surface (given that the surface currents $\mathbf{K} = \mathbf{n} \times \nabla{\Phi}$ flow along these contours, as $\mathbf{K}\cdot \nabla\Phi=0$), given in the surface angle coordinates $(\tp,\zp)$
    \item Find the points in real space geometry corresponding to these coils, by evaluating the winding surface geometry $R^b(\tp,\zp)$ and $Z^b(\tp,\zp)$.
    \item Determining the current to assign to each coil
\end{enumerate}

The following sections will detail the approach to this process and a demonstration case to show the results.

\subsubsection{Demonstration Case}
\begin{figure}
    \centering
    \includegraphics[keepaspectratio, width=3in]{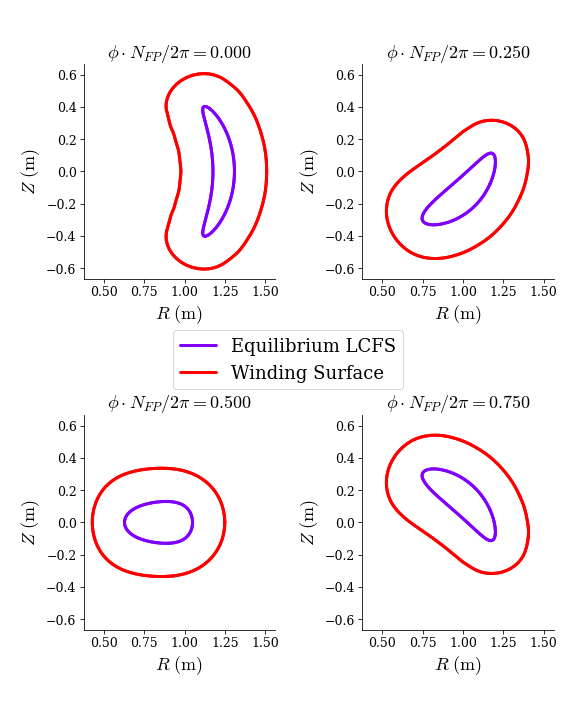}
    \caption{Precise QA equilibrium last closed flux surface (LCFS) and the winding surface plotted at four different toroidal planes, which will be used to demonstrate the coil-cutting procedure.}
    \label{fig:eq_and_wind_surf}
\end{figure}

For this section, coils will be found for the (vacuum) precise QA equilibrium of Landreman and Paul \citep{landreman_magnetic_2022}. The specific equilibrium used was obtained with \texttt{DESC} and is available at the \texttt{DESC} repository \citep{desc_git}. The winding surface used is a constant-offset surface from the last closed flux surface of the equilibrium, with an offset of 0.2 meters, found using the constant offset algorithm detailed by Landreman \citep{landreman_improved_2017} which has been implemented in the \texttt{DESC} code. The last-closed flux surface and the winding surface are plotted at different toroidal planes in Figure \ref{fig:eq_and_wind_surf}. Then, the \texttt{REGCOIL} algorithm implemented in \texttt{DESC} was used to find surface currents minimizing the normal field on the equilibrium boundary, at two different current helicities: $\bar{q}=1, ~p=0$, for modular and $\bar{q}=-1,~p=1$ for helical. A regularization parameter of $\lambda = 10^{-20}$ was used for the calculation. This then yields two different current potentials that will be shown in the following sections to demonstrate the coil-cutting procedure, which is implemented and available in the \texttt{DESC} code\citep{desc_git}.

\subsubsection{Finding Contours on Winding Surface}
The first step is typically done by using the marching squares algorithm (available in common libraries, such as the one implemented in the \texttt{matplotlib} Python library \citep{matplotlib_2007}) to, given the current potential $\Phi$ evaluated on a grid in $(\tp,\zp)$ on the surface, find contours as curves in $(\theta', \zeta')$ on the surface along which the current potential is constant. These can then be translated into real-space points by evaluating the winding surface $R(\theta', \zeta'),~Z(\theta', \zeta')$ at the contours found. If a discrete symmetry is present, then the surface current must only be discretized along the domain $\zp \in [0,2\pi/N_{FP}]$, as the discrete symmetry implies that the entire coilset is composed of the unique coils in this toroidal angle range, just repeated toroidally $N_{FP}-1$ times. Then, if the coils sought are modular coils, the coils may simply be rotated around the cylindrical Z-axis toroidally $N_{FP}-1$ times by $\phi=2\pi/N_{FP}$ each time to form the complete coilset. A diagram of the coil-cutting process for modular coils is shown in Figure \ref{fig:modular_coil_cutting}. Note that if the coils are also expected to be stellarator symmetric, then one only needs to define coils over a half-field-period, and then one may reflect the coils over the symmetry plane to obtain the rest of the coils in that period, then rotate them toroidally around as stated previously to obtain the full coilset, flipping the currents of the reflected coils in order to preserve the stellarator symmetry of the generated magnetic field, $B_R(\rho,-\theta,-\zeta) =- B_R(\rho,\theta,\zeta)$, $B_{\phi}(\rho,-\theta,-\zeta) =- B_{phi}(\rho,\theta,\zeta)$, $B_Z(\rho,-\theta,-\zeta) =B_Z(\rho,\theta,\zeta)$ \citep{strickler_designing_2002, dewar_stellarator_1998}.
\par
If instead the contour topologies are helical, then one must rigidly rotate the unique section (unique section being from $\zp=0$ to $\zp=2\pi/N_{FP}$) of the coil around toroidally. In this case, say a given helical contour's coordinates is given by a set of $m$ points $(\tp_i, \zp_i)$ with $i$ being the index for each coordinate, with $\zp_1=0$ and $\zp_m=2\pi/N_{FP}$, as shown in the bolded purple curve in Figure \ref{fig:helical_coil_cutting}. To then create the full helical coil, the contour's $\zp_i$ are simply repeated $N_{FP}-1$ times with a $2\pi/N_{FP}$ toroidal rotation added each time, so that the set of toroidal angle points along the full curve is $\{\zeta_i + k2\pi/N_{FP}\}~k=0...N_{FP}-1$. The corresponding $\theta_i$ points are then $\{\theta_i + k2\pi\bar{q}\}~k=0...N_{FP}-1$ where $\bar{q}$ is the number of times a single contour rotates poloidally per toroidal field period. In practice, when finding the contours the domain in $\tp$ plotted must be larger than just $[0,2\pi]$, as one must find the whole helical contour from $\zp\in[0,2\pi/N_{FP})$, and depending on the $\bar{q}$ and topology of the coilset, the domain in $\tp$ must be increased. For instance, the helical contour that corresponds to $\Phi(\tp=0, \zp=0)$ will end at $(-2\pi p\bar{q}, \zp=2\pi/N_{FP})$, and if the contour curves are found starting from $\zp=0$, the subsequent contours (subsequent meaning beyond the first coil, e.g. if we want 4 helical coils we will start contours at initial $\tp$ values of $\tp=[0,\pi/2,\pi,3\pi/2]$) will end at poloidal angles larger than $-2\pi\bar{q}$ (in the case in of the previous paranthetical, the last coil will end at $(\tp,\zp) =(-2\pi\bar{q} + 3\pi/2, 2\pi/N_{FP})$). Here the negative sign comes from the fact that if the contours are always considered starting at $\zp=0$ and ending at $\zp=2\pi/N_{FP}$, then the direction in which $\theta$ changes along the contour is the negative of $\bar{q}$ (The same statement is given mathematically in Eq. \eqref{eq:helicity_def}, and can be visualized by looking at the topologies of the contours in Figure \ref{fig:current_helicity}).  A rule of thumb which has been found to work well is to use a domain $\tp \in [sign(\bar{q})2\pi,-sign(\bar{q})(2\pi (|\bar{q}|+1))] $, with the $sign(\bar{q})$ accounting for the fact that whether coil contours are increasing or decreasing in $\tp$ as $\zp$ is increased depends on the sign of $\bar{q}$.

\subsubsection{Finding Real-Space Geometry of the Coils}
For the second step, the above $(\tp_i,\zp_i)$ are simply plugged into the winding surface geometry (typically given as a Fourier series in $(\tp,\zp)$ for the cylindrical radius $R$ and the vertical position $Z$), to get a set of points in real space $(R_i,\phi_i,Z_i)$ describing the position of points along the coil. These can then be fit with a spline or a Fourier series in some curve parameter like arc length (usually after converting the position to Cartesian coordinates, e.g. to use the \texttt{SplineXYZCoil} and \texttt{FourierXYZCoil} classes in \texttt{DESC}), in order to get a continuous representation of the coil.
\par
\begin{figure}
  \begin{subfigure}[t]{0.85\textwidth}
    \includegraphics[height=2.5in,keepaspectratio]{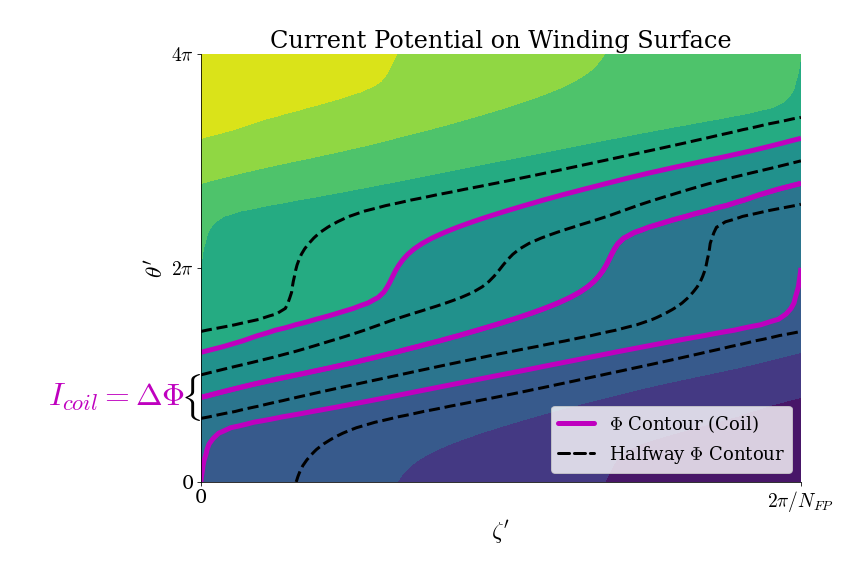}
    \caption{Helical Current Potential Contours}
    \label{fig:helical_coil_cutting}
  \end{subfigure}
  \begin{subfigure}[t]{0.88\textwidth}
    \centering
    \includegraphics[height=2.5in,keepaspectratio]{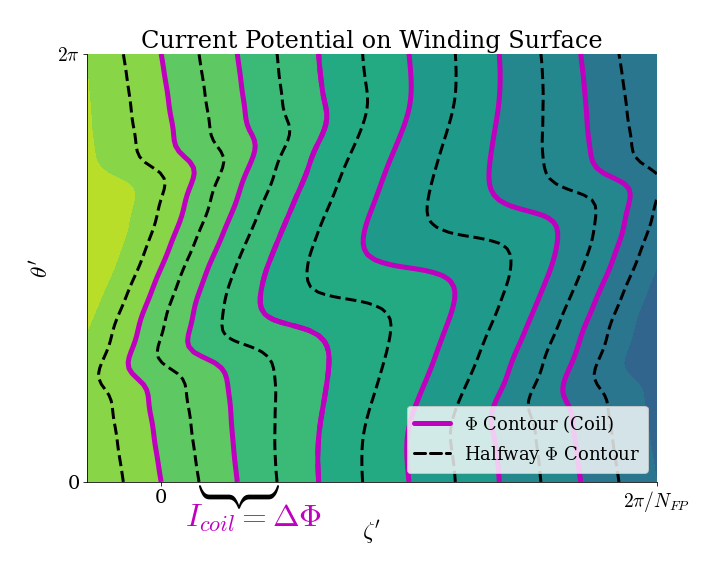}
    \caption{Modular Current Potential Contours}
    \label{fig:modular_coil_cutting}
  \end{subfigure}
  \caption{Schematic of cutting coils from (a) a current potential with $I=G$ (with $\bar{q}=-1$) cut into 3 helical coils, and (b) a current potential with $I=0$ cut into $6N_{FP}$ modular coils (6 coils per field period).  In solid purple lines are candidate contours for coils, and in dashed black are the contours halfway in $\Phi$ value between adjacent coil contours.}
  \label{fig:coil_cutting}
\end{figure}

For the last step, the current that each coil is assigned can be related to the difference in the current potential between that coil and its neighboring contours, as shown in Section \ref{sec:relations}. This information can be used to assign current by assigning a given coil the net current flowing between a region halfway between that coil and its neighboring coils.
A reasonable choice of the current potential contours is to simply choose them to be equally spaced in $\Phi$ from $[0,I)$ (helical) or $[0,G/N_{FP})$ for modular ($N_{coils}/N_{FP}$ equally spaced contours, with the coils then repeated $N_{FP}$ times to form the full $N_{coils}$ coils), and then the current per coil is equal, since the difference in $\Phi$ between all contours is equal by construction. This current is $I/N_{coils}$ for helical coils or $G/N_{coils}$ for modular coils (where here $N_{coils}$ is the total number of coils, not just the number of coils in a single field period). Note that the modular coil currents add up to $G$ as expected, and the helical coil currents linking the plasma poloidally also add up to $G$, but in general each helical coil may wrap around the plasma multiple times poloidally per toroidal transit, the relation is not simply $G = I_{coil}N_{coils}$ like for modular coils but rather $G = I_{coil}N_{coils}(winding~number)$ where $winding~number$, as previously defined, is the number of poloidal turns a coil makes as it makes a single toroidal transit. This is the choice of current that has been used conventionally \citep{pomphrey_innovations_2001}, and can be thought of physically as choosing to assign each coil all of the current that is flowing between the region around the coil extending halfway to each neighboring coil, as illustrated in Figure \ref{fig:coil_cutting}. 
\par
\begin{figure}
  \begin{subfigure}[t]{0.475\textwidth}
    \includegraphics[width=1.05\textwidth,keepaspectratio]{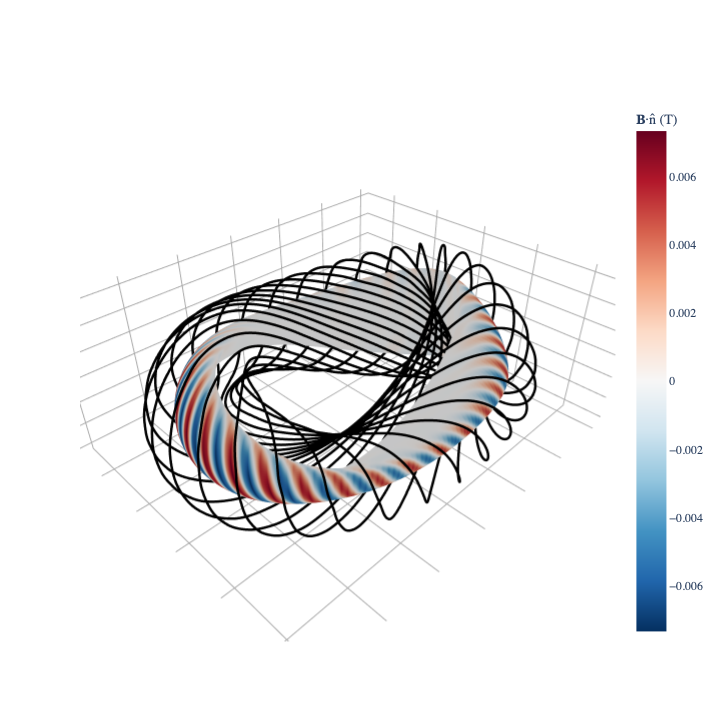}
    \caption{Helical Coilset}
    \label{fig:helical_coils}
  \end{subfigure}
  \begin{subfigure}[t]{0.475\textwidth}
    \includegraphics[width=1.05\textwidth,keepaspectratio]{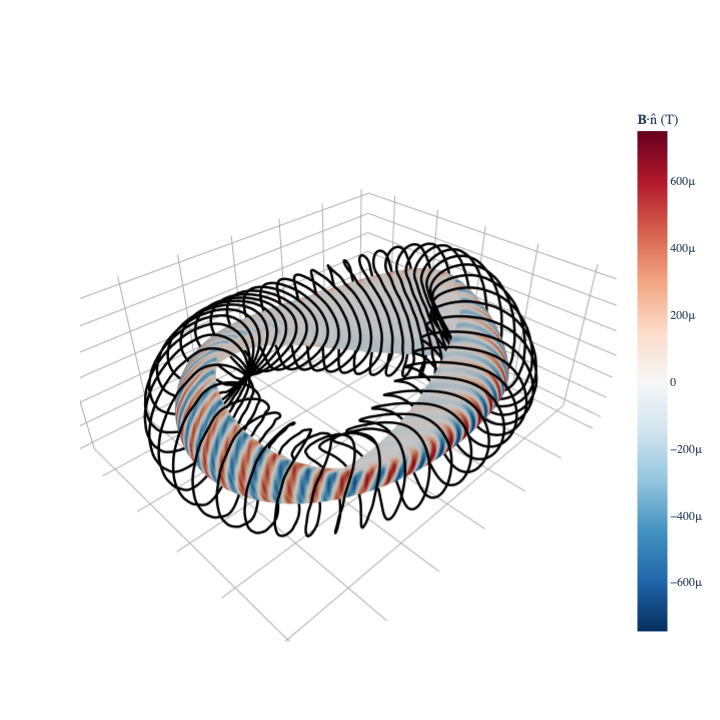}
    \caption{Modular Coilset}
    \label{fig:modular_coils}
  \end{subfigure}
  \caption{Three-dimensional figures of (a) a helical coilset with 16 coils and (b) a modular coilset with 12 coils per field period (24 coils total), plotted over the equilibrium (with the normal field $\mathbf{B}\cdot \mathbf{n}$ plotted on its surface). Each coilset was cut from the current potentials shown in Section \ref{sec:cutting_helical_coils}. }
\end{figure}
\begin{figure}
  \begin{subfigure}[t]{0.45\textwidth}
    \includegraphics[width=0.9\textwidth,keepaspectratio]{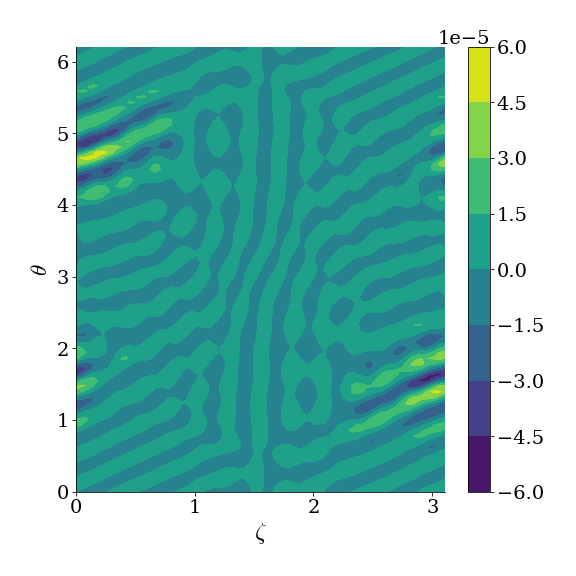}
    \caption{Normal field from Helical Surface Current}
    \label{fig:helical_surf_current_Bn}
  \end{subfigure}
  \begin{subfigure}[t]{0.45\textwidth}
    \includegraphics[width=0.9\textwidth,keepaspectratio]{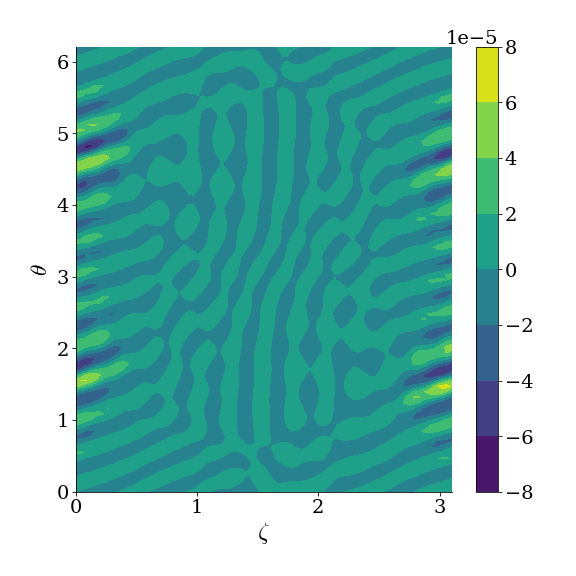}
    \caption{Normal field from Modular Surface Current}
    \label{fig:modular_surf_current_Bn}
  \end{subfigure}
  \caption{The normal field error ($|\mathbf{B}\cdot\mathbf{n}|$) plotted over the equilibrium surface for the (a) helical surface current and (b) modular surface current.}
    \label{fig:Bn_surface_currents}
\end{figure}
\begin{figure}
  \begin{subfigure}[t]{0.475\textwidth}
    \includegraphics[width=1.05\textwidth,keepaspectratio]{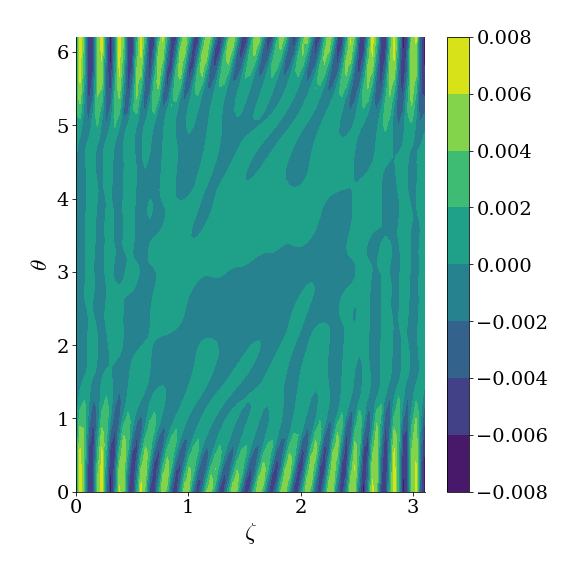}
    \caption{Normal field from Helical Coilset}
    \label{fig:helical_coils_Bn}
  \end{subfigure}
  \begin{subfigure}[t]{0.475\textwidth}
    \includegraphics[width=1.05\textwidth,keepaspectratio]{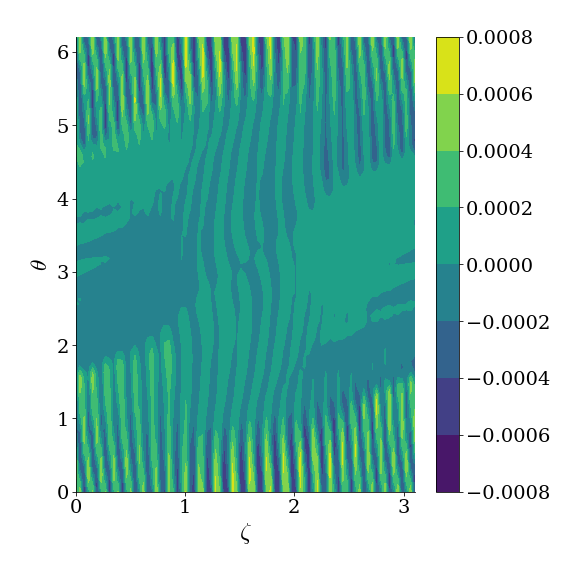}
    \caption{Normal field from Modular Coilset}
    \label{fig:modular_coils_Bn}
  \end{subfigure}
  \caption{The normal field error ($|\mathbf{B}\cdot\mathbf{n}|$) plotted over the equilibrium surface for the (a) helical coilset with 16 coils and (b) modular coilset with 12 coils per field period (24 coils total). Note the increase in normal field versus Figure \ref{fig:Bn_surface_currents}, as after the coil discretization the current must be constant along the coils, while for the surface current the currents may vary along each constant current potential contour.}
    \label{fig:Bn_coils}
\end{figure}
\begin{figure}
  \begin{subfigure}[t]{0.5\textwidth}
    \includegraphics[width=1.25\textwidth,keepaspectratio]{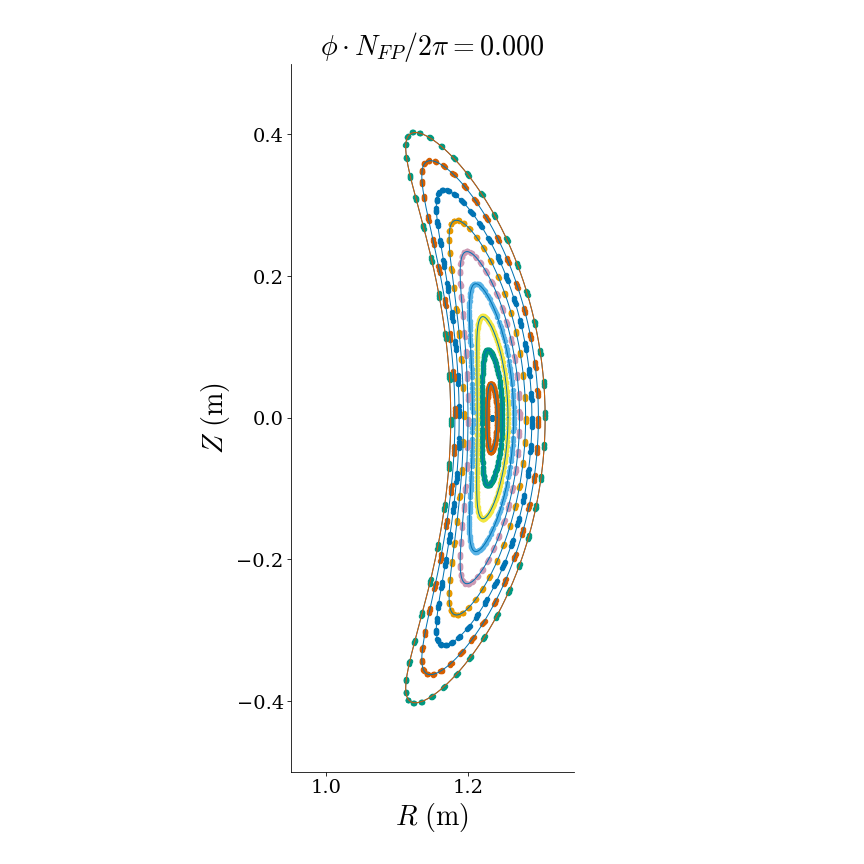}
    \caption{Poincare Plot from Helical Coilset}
    \label{fig:helical_coils_trace}
  \end{subfigure}
  \begin{subfigure}[t]{0.5\textwidth}
    \includegraphics[width=1.25\textwidth,keepaspectratio]{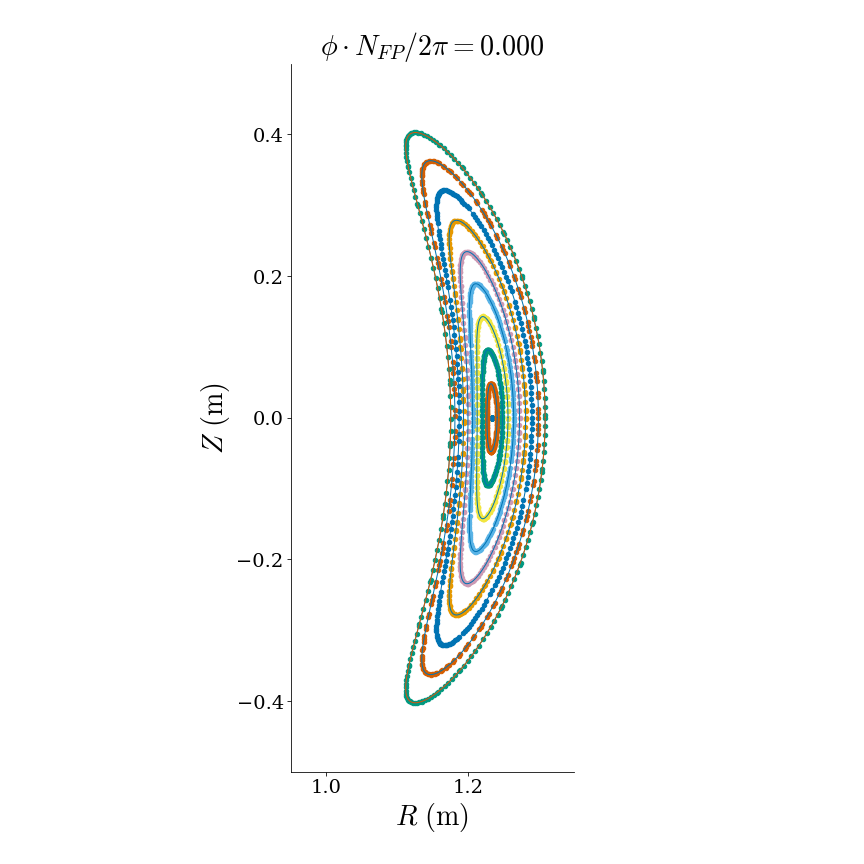}
    \caption{Poincare Plot from Modular Coilset}
    \label{fig:modular_coils_trace}
  \end{subfigure}
  \caption{Poincare plot of the magnetic field lines from the coilsets at the $\phi=0$ plane, plotted over the equilibrium flux surfaces for the (a) helical coilset with 16 coils and (b) modular coilset with 12 coils per field period (24 coils total).}
    \label{fig:field_trace_coils}
\end{figure}

The final products of the workflow for a modular coilset with 12 coils per field period and a helical coilset ($\bar{q}=-1$) with 16 coils are shown in Figures \ref{fig:helical_coils} and  \ref{fig:modular_coils}. Plots of the normal field on the plasma boundary from the surface currents are shown in Figure \ref{fig:Bn_surface_currents}, and the normal field plots from the resulting coils are shown in Figure \ref{fig:Bn_coils} for each coilset. Poincare plots from field line tracing at the $\phi=0$ toroidal plane are shown in Figure \ref{fig:field_trace_coils} for each coilset, showing flux surfaces in the magnetic field generated by the coils, in good agreement with the equilibrium's flux surfaces.
\par
In a practical application, at this point turning to a filamentary coil optimization (with the equal current constraint relaxed \cite{jr_use_2001}) is warranted, as the coils can be modified in an attempt correct the field errors introduced through the discretization errors of (i) taking only a discrete number of contours, when the surface current itself is a continuum of current-carrying contours, (ii) assuming equal current along the current potential contours, and (iii) restricting the surface currents to lie on a single, fixed winding surface.

% just add one more result of varying helicity and maybe both winding surf and current opt then will be totally done 
% \section{Investigating Different Current Helicities}
% \textcolor{red}{actually I might leave this for later and not include here}
% The user is free to specify the net toroidal current $I$ of a \texttt{REGCOIL} solution, so it is of interest to investigate how the solution changes with this choice. Merkel states that the magnetic field in the domain inside of the winding surface depends solely on the choice of the net poloidal current $G$, and that $I$ simply defines a family of solutions which should yield similar fields. To inspect the ramifications of this statement in practice, we will run the \texttt{REGCOIL} algorithm by scanning over the regularization term $\lambda$ for a fixed equilibrium and winding surface, yielding a curve of $\chi_B^2$ versus $\lambda$. We will perform this scan for different current helicities $p$ for different equilibria to see if there is any effect on the solution. The same grids for evaluation and source will be used for all cases \textcolor{red}{write the grid I use}.

\section{Conclusion and Future Work}

In this work, the basic theory and some derivations relating to the surface current method of the stage-two coil optimization were presented. The accompanying coil discretization algorithm was described in-depth, and an example was shown. The \texttt{REGCOIL} algorithm shown in this paper was implemented in the \texttt{DESC} equilibrium code, and is capable of running on both CPUs and GPUs.
\par
Some possible improvements to the coil-cutting algorithm include removing the reliance on the marching squares algorithm to find the contours, which requires one to plot the current potential values on a discrete grid in the correct domain of interest in order to be successful. This method can fail when the contours are too strongly shaped and thus do not all lie in the domain being plotted. Treating the contour as $\tp(\zp)$ (or $\zp(\tp)$) and root-finding in the independent variable, or solving an ODE for the contour coordinates $\tp(s)$ and $\zp(s)$ (with $s$ being an arclength-like parameter) are two possible, more accurate alternatives, but they may be slower as they require the contour to be constructed sequentially (in the case of the ODE), or the contour may double back on itself and thus not be a single-valued function of $\zp$ or $\tp$ (causing root-finding to fail). A third, promising alternative is to construct a parametrization of a curve which lies on the winding surface with the desired topology, and then perform an optimization to minimize $(\Phi(\tp_c,\zp_c) - \Phi_{target})^2$, where $\tp_c$ and $\zp_c$ are the coordinates of the curve. The curve parametrization detailed by Strickler and used in \texttt{COILOPT} \citep{strickler_designing_2002}, for example, could be used, where the curve topology is determined by the secular terms in Eq. 2 of the aforementioned reference. With this method of cutting coils, there would be no need for separate discretization algorithms, and a single algorithm could handle modular, helical, saddle or even umbilic-type coils. In the future, this coil cutting algorithm is planned to be implemented in the \texttt{DESC} code.
\par
The quadratic flux and surface current regularization cost functions were also implemented as objective functions in \texttt{DESC}, allowing for their use in conventional optimization problems. One possible use case could be optimization of the winding surface and the surface current. This has been done in the past using adjoint methods in the \texttt{REGCOIL} code \citep{paul_adjoint_2018}, in \texttt{DESC} this could be expanded to include any objective, with derivatives obtained by automatic differentiation. The quadratic flux and surface current regularization can also be included in an optimization of the equilibrium along with physics objectives like quasisymmetry to allow for single-stage optimization using surface currents. Using surface currents may hold advantages over single-stage optimization with filamentary coils, as the outcome of filamentary coil optimization is highly dependent on the number of coils and their initial positions and currents. The surface current approach, on the other hand, would depend mainly on the initial geometries of the winding surface and equilibrium, and the current potential resolution, with no choice needing to be made for number of coils during the optimization.

\section{Data Availability}
All figures and results in the paper were created using the \texttt{DESC} code, and the relevant scripts and outputs are available upon request to the author.

\section{Acknowledgements}
One of the authors (D.P.) would like to acknowledge helpful discussions with John Kappel and Frank Fu on the mathematics behind the coil discretization algorithm.

\section{Funding}
% the PPPL stell foundry grant 
This work is funded through the SciDAC program by the US Department of Energy, Office of Fusion Energy Science and Office of Advanced Scientific Computing Research under contract No. DE-AC02-09CH11466, as well as DE-SC0022005. The United States Government retains a non-exclusive, paid-up, irrevocable, world-wide license to publish or reproduce the published form of this manuscript, or allow others to do so, for United States Government purposes. This work was also funded by nT-Tao Ltd. and the Israeli innovation authority grant number 80677 and under the contract number NT-Tao-10015925

\newpage
%%%%%%%%%%%%%%%%%%%%%%%%%%%%%%%%%%%%%%%%%%%%%%%%%%%%%%%%%%%%%%%%
%%%%%%%%%%%%%%%%%%%%%%%  Appendix  %%%%%%%%%%%%%%%%%%%%%%%%%%%%%
%%%%%%%%%%%%%%%%%%%%%%%%%%%%%%%%%%%%%%%%%%%%%%%%%%%%%%%%%%%%%%%%
\appendix

\section{Determining the Current Flowing In Between Current Potential Contours}\label{app:G_derivation}

In this appendix the statements given in Section \ref{sec:relations} are derived in more detail. First, the surface current density contravariant components are found starting from:

\begin{equation}
    \bm{K} =\bm{n} \times  \nabla \Phi ,
\end{equation}

where $\bm{n} =\frac{\bm{N}}{|\bm{N}|}= \frac{\bm{e}_{\tp} \times \bm{e}_{\zp}}{|\bm{e}_{\tp} \times \bm{e}_{\zp}|}$ is the unit surface normal vector to the toroidal winding surface. Using the identities $\bm{e}_{\tp} = \bm{N} \times \nabla \zp$ and $\bm{e}_{\zp} = \bm{N} \times \nabla \tp$:

\begin{align}
    \bm{K} &=\bm{n} \times  \nabla \Phi = \bm{n} \times  \nabla \left( \Phi_{SV}(\theta',\zeta') + \frac{G\zeta'}{2\pi} + \frac{I\theta'}{2\pi} \right) \\
    &=  \left(\p{\Phi_{SV}}{\zeta'} + \frac{G}{2\pi}\right)\bm{n} \times \nabla \zeta' + \left(\p{\Phi_{SV}}{\theta'} + \frac{I}{2\pi}\right)\bm{n} \times \nabla \theta'   \\
    &=  \frac{1}{|\bm{N}|}\left( - \left(\p{\Phi_{SV}}{\zeta'} + \frac{G}{2\pi}\right)\bm{e}_{\theta'} + \left(\p{\Phi_{SV}}{\theta'} + \frac{I}{2\pi}\right)\bm{e}_{\zeta'}\right) \\
    &=  \frac{1}{|\bm{N}|}\left( - \p{\Phi}{\zeta'}\bm{e}_{\theta'} + \p{\Phi}{\theta'}\bm{e}_{\zeta'}\right) \\
    &= K^{\theta} \bm{e}_{\theta'} + K^{\zeta} \bm{e}_{\zeta'}
\end{align}

where
\begin{align}
    K^\theta &= -\frac{1}{|\bm{N}|}\left(\p{\Phi_{SV}}{\zeta'} + \frac{G}{2\pi}\right)\\
    K^\zeta &= \frac{1}{|\bm{N}|}\left(\p{\Phi_{SV}}{\theta'} + \frac{I}{2\pi}\right)\\
\end{align}
are the contravariant components of the surface current density. It should be noted that the contravariant basis vectors $\nabla\tp$ and $\nabla\zp$ are not unique, as the angular coordinates are defined only on the winding surface, so these vectors depend on the choice of how one extends the angular coordinates off-surface. One could choose, for instance, to define them such that $\bm{n} \cdot \nabla\tp=\bm{n} \cdot \nabla\zp=0$.  The results in this appendix, however, are independent of the particular choice  of the extension off-surface. For the purposes of plotting, the contravariant basis vectors are defined by taking the initial surfaces and scaling them down to their centroid, thus defining interior surfaces with the same angle definitions as the boundary. This thus defines the contravariant basis vectors shown in the plots in this appendix.
\par
The net current linking the surface and the plasma poloidally (i.e. the net poloidal current on the surface, $I_{pol}$) is related to the magnetic field inside as:

\begin{align}
   I_{pol} = -G &= \frac{1}{\mu_0}N_{FP}\int_{0}^{2\pi/N_{FP}} B_{\zeta} d\zeta\\
\end{align}

\begin{figure}
    \centering
    \includegraphics[keepaspectratio, width=3in]{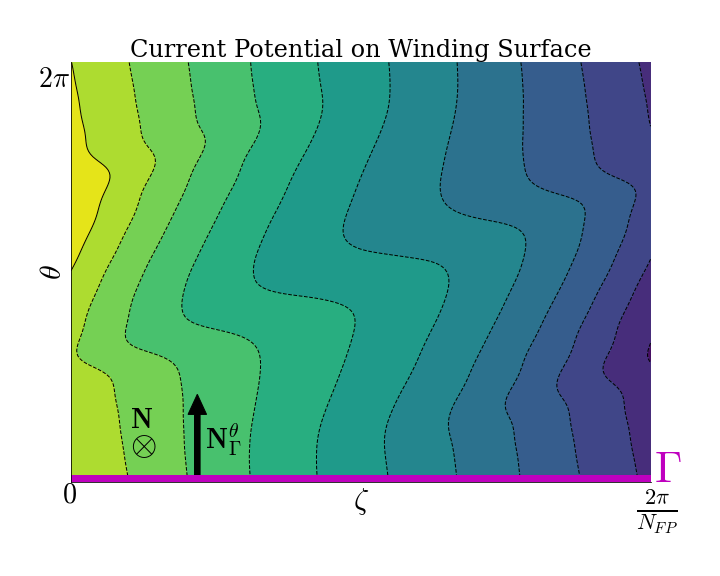}
    \caption{Schematic of a current potential on a winding surface, with the normal vector $\bm{N}_{\Gamma}^{\theta}$ to the constant $\theta$ curve $\Gamma$ and the surface normal vector $\bm{N}$ to the winding surface.}
    \label{fig:net_pol_current}
\end{figure}

\begin{figure}
  \begin{subfigure}[t]{0.475\textwidth}
\includegraphics[width=1.25\textwidth,keepaspectratio]{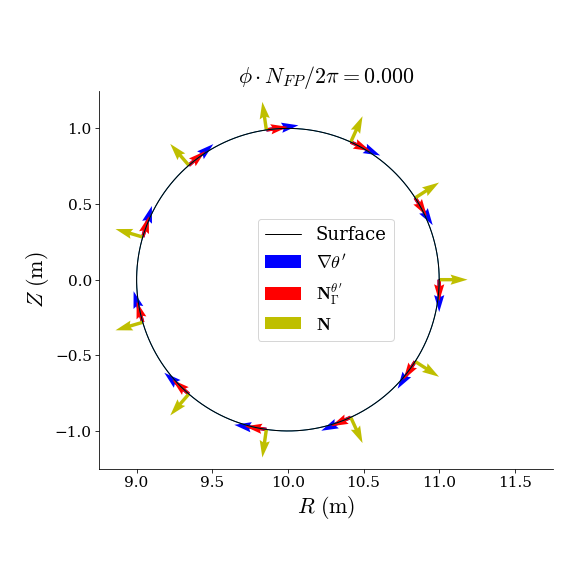}
    \caption{Circular axisymmetric torus}
  \end{subfigure}
  \begin{subfigure}[t]{0.475\textwidth}
\includegraphics[width=1.25\textwidth,keepaspectratio]{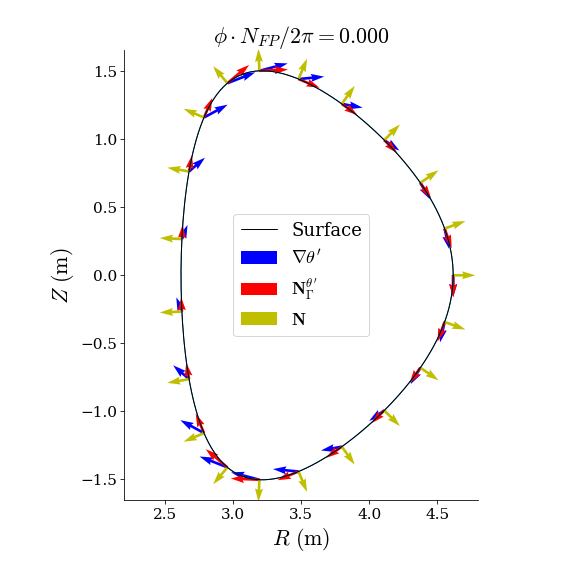}
    \caption{D-shaped axisymmetric torus}
  \end{subfigure}
  \caption{Diagram of the different vectors defined in Section \ref{sec:relations} for (a) a circular, axisymmetric toroidal surface and (b) an axisymmetric D-shaped toroidal surface. It can be seen that with the stated choice of $\nabla \tp$ there is no distinction between $\nabla\tp$ and $\bm{N}_{\Gamma}^{\tp}$ for the simplest toroidal shape, but differences emerge when non-trivial shaping is considered (as is the case in every stellarator and most tokamaks)}
  \label{fig:diagram-vectors}

\end{figure}

The equality of $I_{pol}$ to the negative of the parameter $G$, can be shown as follows. First, some geometric quantities:

\begin{align}
    \bm{N} &= \bm{e}_{\theta} \times \bm{e}_{\zeta}\\
    \bm{n} &= \frac{\bm{N}}{|\bm{N}|}\\
    \bm{N}_{\Gamma}^{\theta} &= \nabla \theta - \frac{(\bm{N}\cdot \nabla \theta)\bm{N}}{|\bm{N}|^2}\\
    \bm{n}_{\Gamma}^{\theta} &= \frac{\bm{N}_{\Gamma}^{\theta}}{|\bm{N}_{\Gamma}^{\theta}|}   , 
\end{align}

where $\bm{n}$ is the unit normal to the winding surface and $\bm{n}_{\Gamma}^{\theta}$ is the normal vector to a constant $\theta$ curve (denoted $\Gamma$) projected onto the surface, normalized, as shown in Figure \ref{fig:net_pol_current}, with the different vector directions shown in Figure \ref{fig:diagram-vectors} for a circular axisymmetric torus and an axisymmetric D-shaped toroidal surface (an axisymmetric surface is chosen only to simplify the projections to be in two dimensions, for demonstration purposes).
    
To relate $I_{pol}$ to $G$, we will write the expression for the net poloidal current $I_{pol}$ as the surface current density $\bm{K}$ dotted with the unit normal to the curve $\Gamma$, integrated along that curve:
\begin{align}
    I_{pol} &= \int_{\Gamma} \bm{K} \cdot \bm{n}_{\Gamma}^{\theta} dl\\
    &= N_{FP}\int_{0}^{2\pi/N_{FP}}  \bm{K} \cdot \bm{n}_{\Gamma}^{\theta} \sqrt{g_{\zeta\zeta}}d\zeta\\
    &= N_{FP}\int_{0}^{2\pi/N_{FP}} \frac{(K^{\theta} \bm{e}_{\theta} + K^{\zeta} \bm{e}_{\zeta}) \cdot \bm{n}_{\Gamma}^{\theta} \sqrt{g_{\zeta\zeta}}}{|\bm{N}|} d\zeta\\
    &\text{Use the fact that }\bm{e}_{\zeta} \cdot\bm{n}_{\Gamma}^{\theta} = 0 \\
    &= N_{FP}\int_{0}^{2\pi/N_{FP}} (K^{\theta} \bm{e}_{\theta}) \cdot \bm{n}_{\Gamma}^{\theta} \sqrt{g_{\zeta\zeta}} d\zeta\\
    &= -N_{FP}\int_{0}^{2\pi/N_{FP}} \frac{ \left(\p{\Phi_{SV}}{\zeta'} + \frac{G}{2\pi}\right) \bm{e}_{\theta} \cdot \bm{n}_{\Gamma}^{\theta} \sqrt{g_{\zeta\zeta}}}{|\bm{N}|} d\zeta,\\
\end{align}

Let's focus for now on the quantity inside the integrand (defined as $\Xi$) multiplying the paranthetical $\left(\p{\Phi_{SV}}{\zeta'} + \frac{G}{2\pi}\right)$, and show that it reduces to unity:

\begin{align}
    \frac{\bm{e}_{\theta} \cdot \bm{n}_{\Gamma}^{\theta} \sqrt{g_{\zeta\zeta}}}{|\bm{N}|} &= \frac{\sqrt{g_{\zeta\zeta}}}{|\bm{N}||\bm{N}_{\Gamma}^{\theta}|}\left(\cancel{\bm{e}_{\theta}\cdot \nabla \theta}^{~1} - \frac{(\bm{N} - \nabla \theta)\cancel{(\bm{e}_{\theta}\cdot \bm{N})}^{~0}}{|\bm{N}|^2} \right)\\
    &= \frac{\sqrt{g_{\zeta\zeta}}}{|\bm{N}||\bm{N}_{\Gamma}^{\theta}|}\\
    &= \frac{\sqrt{g_{\zeta\zeta}}}{\sqrt{\bm{N}\cdot\bm{N}} \sqrt{\bm{N}_{\Gamma}^{\theta} \cdot \bm{N}_{\Gamma}^{\theta}}}\\
   \Xi &= \sqrt{\frac{g_{\zeta\zeta}}{(\bm{N}\cdot\bm{N}) (\bm{N}_{\Gamma}^{\theta} \cdot \bm{N}_{\Gamma}^{\theta})}},
\end{align}

Consider the square of this quantity:
\begin{align}
    \Xi^2 &= \frac{g_{\zeta\zeta}}{(\bm{N}\cdot\bm{N}) (\bm{N}_{\Gamma}^{\theta} \cdot \bm{N}_{\Gamma}^{\theta})}\\
    &= \frac{g_{\zeta\zeta}}{(\bm{N}\cdot\bm{N})\left( g^{\theta\theta} - 2 \frac{(\bm{N}\cdot \nabla \theta)(\nabla \theta \cdot \bm{N})}{|\bm{N}|^2} + \frac{(\bm{N}\cdot \nabla \theta)^2(\bm{N}\cdot \bm{N})}{|\bm{N}|^4} \right)}\\
    &= \frac{g_{\zeta\zeta}}{ g^{\theta\theta}(\bm{N}\cdot\bm{N}) - 2 (\bm{N}\cdot \nabla \theta)^2 + (\bm{N}\cdot \nabla \theta)^2 }\\
    &= \frac{g_{\zeta\zeta}}{ g^{\theta\theta}(\bm{N}\cdot\bm{N}) -  (\bm{N}\cdot \nabla \theta)^2  }\\
    &= \frac{\bm{e}_{\zeta} \cdot \bm{e}_{\zeta}}{ (\nabla\theta\cdot\nabla\theta)(\bm{N}\cdot\bm{N}) -  (\bm{N}\cdot \nabla \theta)^2  },\\
\end{align}

Now we will remind ourselves that $\bm{e}_{\zeta} = \bm{N} \times \nabla \theta$ and use a vector identity $(\bm{A} \times \bm{B})\cdot(\bm{C} \times \bm{D}) = (\bm{A} \cdot \bm{C})(\bm{B}\cdot \bm{D}) - (\bm{B}\cdot \bm{C})(\bm{A} \cdot \bm{D})$ to find that:
\begin{align}
    \bm{e}_{\zeta} \cdot \bm{e}_{\zeta} &= (\bm{N} \times \nabla \theta) \cdot (\bm{N} \times \nabla \theta)\\
    &= (\bm{N}\cdot \bm{N})(\nabla \theta \cdot \nabla \theta) - (\bm{N} \cdot\nabla\theta)(\bm{N} \cdot\nabla\theta)\\
    &= (\nabla \theta \cdot \nabla \theta)(\bm{N}\cdot \bm{N}) - (\bm{N} \cdot\nabla\theta)^2,
\end{align}

and therefore the square of $\Xi$ is equal to $1$, so $\Xi$ itself is equal to $1$ (positive $1$, since the integrand is a square root and we only consider real numbers here), everywhere in the integration interval. Returning to the original integral after setting the aforementioned term to unity:

\begin{align}
    I_{pol} &=-N_{FP}\int_{0}^{2\pi/N_{FP}}  \left(\p{\Phi_{SV}}{\zeta} + \frac{G}{2\pi}\right) d\zeta\\
    &= \underbrace{-\cancel{N_{FP}\int_{0}^{2\pi/N_{FP}}  \p{\Phi_{SV}}{\zeta}d\zeta}}_{0 \text{ b/c } \Phi_{SV} \text{ is periodic in $\zeta$ with period $2\pi/N_{FP}$}}  -N_{FP}\int_{0}^{2\pi/N_{FP}} \frac{G}{2\pi} d\zeta\\
    &= -\frac{N_{FP}}{2\pi} \frac{2\pi}{N_{FP}}G\\
    \Aboxed{I_{pol}&= -G},
\end{align}

Thus, we have shown (for an arbitrarily shaped toroidal surface) that the parameter $G$ is the negative of the net poloidal current flowing due to the surface current on the winding surface. 
\par
Further, if the curve $\Gamma$ were not taken across the entire period in $\zeta$ but instead between two points $\zeta_0$ and $\zeta_1$, we find that the current passing between two constant current potential contours is simply equal to the difference in the current potential $\Phi$ of those two contours:

\begin{align}
    I_{\Delta \zeta} &= \int_{\zeta_0}^{\zeta_1} \left(\p{\Phi_{SV}}{\zeta} + \frac{G}{2\pi}\right) d\zeta\\
    &= \int_{\zeta_0}^{\zeta_1} \p{\Phi}{\zeta} d\zeta\\
    I_{\Delta \zeta} &= \Phi(\theta,\zeta_1) - \Phi(\theta,\zeta_1)\\
    \Aboxed{I_{\Delta \zeta} &= \Phi_1 - \Phi_0},
\end{align}

where $I_{\Delta \zeta}$ is the current flowing between the two contours and $\Phi_1$ and $\Phi_0$ are the value of the current potential along the two given contours. Nothing is assumed about the existence of windowpane contours in the current potential, so the result holds in that case as well. While the contours shown are not windowpane coils, the same logic applies to windowpane contours, where the difference in the current potential between the center of the windowpane and the edge is equal to the total current circulating within that windowpane.

An analogous derivation can show that the parameter $I$ is the (positive) net toroidal current flowing around the winding surface due to the surface current $I_{tor}$,
\begin{equation}
    \boxed{I_{tor} = I}
\end{equation}

These results were first stated, without derivation, in a work by Boozer \citep{boozer_optimization_2000}, but to the author's knowledge this is the first time the derivation is presented.

\bibliographystyle{jpp}

\bibliography{references}

\end{document}